\documentclass[%
 reprint,
 amsmath,amssymb,
 aps,
pra,
showkeys
]{revtex4-2}

\usepackage{graphicx}
\usepackage{dcolumn}
\usepackage{bm}
\usepackage{multirow}
\usepackage{xcolor}
\usepackage[ruled,linesnumbered, vlined]{algorithm2e}

\begin{document}

\preprint{APS/123-QED}

\title{Distribution-Adaptive Dynamic Shot Optimization for Variational\\Quantum Algorithms}

\author{Youngmin Kim$^{1}$}
\author{Enhyeok Jang$^{1}$}
\author{Hyungseok Kim$^{1}$}
\author{Seungwoo Choi$^{1}$}
\author{Changhun Lee$^{1}$}
\author{Donghwi Kim$^{2}$}
\author{Woomin Kyoung$^{2}$}
\author{Kyujin Shin$^{2}$}
\email{shinkj@hyundai.com}
\author{Won Woo Ro$^{1}$}
\email{wro@yonsei.ac.kr}
\affiliation{$^1$School of Electrical and Electronic Engineering, Yonsei University, Seoul 03722, Republic of Korea \\
$^2$Materials Research \& Engineering Center, Advanced Vehicle Platform Division, Hyundai Motor Company, Uiwang 16082, Republic of Korea}


\begin{abstract}
Variational quantum algorithms (VQAs) have attracted remarkable interest over the past few years because of their potential computational advantages on near-term quantum devices. They leverage a hybrid approach that integrates classical and quantum computing resources to solve high-dimensional problems that are challenging for classical approaches alone. In the training process of variational circuits, constructing an accurate probability distribution for each epoch is not always necessary, creating opportunities to reduce computational costs through shot reduction. However, existing shot-allocation methods that capitalize on this potential often lack adaptive feedback or are tied to specific classical optimizers, which limits their applicability to common VQAs and broader optimization techniques. Our observations indicate that the information entropy of a quantum circuit's output distribution exhibits an approximately exponential relationship with the number of shots needed to achieve a target Hellinger distance. In this work, we propose a distribution-adaptive dynamic shot (DDS) framework that efficiently adjusts the number of shots per iteration in VQAs using the entropy distribution from the prior training epoch. Our results demonstrate that the DDS framework sustains inference accuracy while achieving a $\sim$50\% reduction in average shot count compared to fixed-shot training, and $\sim$60\% higher accuracy than recently proposed tiered shot allocation methods. Furthermore, in noisy simulations that reflect the error rates of actual IBM quantum systems, DDS achieves approximately a $\sim$30\% reduction in the total number of shots compared to the fixed-shot method with minimal degradation in accuracy, and offers about $\sim$70\% higher computational accuracy than tiered shot allocation methods.

\end{abstract}

\maketitle

\section{\label{sec:Introduction}Introduction}

Quantum computing represents a novel computational paradigm, offering the potential for enhanced computational capabilities beyond the reach of classical computers \cite{shor1999polynomial, grover1996fast, arute2019quantum, nielsen2010quantum, farhi2016quantum, jozsa1998quantum}. Variational quantum algorithms (VQAs) \cite{cerezo2021variational, biamonte2017quantum, rebentrost2014quantum, mcclean2016theory, havlivcek2019supervised} are designed to leverage these advantages on noisy intermediate-scale quantum (NISQ) \cite{preskill2018quantum} devices available in the current era. Unlike classical counterparts, quantum computation requires multiple shots to approximate the expected probability distribution of a quantum state, as results cannot be fully observed in a single measurement. While increasing the number of shots can improve the statistical accuracy of results, it also prolongs execution time. Therefore, efficient optimization strategies for shot allocation are crucial to minimize overall computational overhead while maintaining training quality, particularly in iterative VQA training \cite{tilly2022variational}, where circuit evaluations are performed at each training epoch.

Previous research has examined various approaches to reducing the number of shots required to implement quantum circuits without compromising performance. One approach involves improving the optimizer used in the VQA process \cite{ito2023latency, ito2023santaqlaus, luo2022koopman} by focusing on shot allocation based on specific Hamiltonian terms \cite{kubler2020adaptive, arrasmith2020operator, gu2021adaptive} and rearranging these terms to optimize shot distribution \cite{choi2023fluid, yen2023deterministic, choi2022improving}. Additionally, variance-minimization-based shot assignment methods \cite{wecker2015progress, arrasmith2020operator, crawford2021efficient, mniszewski2021reduction, choi2022improving, zhang2023composite, zhu2024optimizing} have been explored to reduce measurement variance. However, these methods often introduce computational overhead, as they require a detailed analysis of Hamiltonian terms or the quantum circuit itself for predictive adjustments. There is potential to efficiently allocate shots in order to minimize total execution time without incurring additional computational costs \cite{zhu2024optimizing, liang2024artificial, kahani2023novel}.

Recently, the tiered shot allocation method \cite{phalak2023shot} has been proposed to enable dynamic shot allocation throughout the optimization process. This method deterministically reduces the overall number of measurement shots by statically defining a reduced shot count. However, reducing the number of shots without considering the specific VQA circuit or variations in the quantum state during training does not guarantee the quality of training parameters, which limits its applicability. To address this limitation, a feedback mechanism is essential to dynamically adjust the shot count for each iteration, ensuring efficient resource utilization while maintaining the consistency and reliability of the VQA.

We have observed a correlation between the entropy of the probability distribution and the number of measurement shots required to guarantee a specific Hellinger distance. This finding suggests that the entropy of the probability distribution can serve as a practical feedback metric for adaptively determining the optimal number of measurement shots. During VQA iterations, probability distributions evolve, ranging from evenly distributed states with high entropy to those concentrated around a few high-probability states with low entropy. Our observations reveal a clear correlation between the characteristics of these distributions and the number of shots required to obtain reliable results. Specifically, evenly distributed probability distributions require a larger number of shots for accurate measurements, whereas distributions concentrated on a small subset of quantum states can be captured with fewer shots.

In this paper, we present the distribution-adaptive dynamic shot (DDS) approach, designed to optimize shot allocation and reduce execution time on quantum computers. Our method dynamically determines the optimal number of shots based on the specific quantum circuit and its probability distribution, leading to reduced shot usage with minimal impact on the overall programmability of the VQA process. By adjusting the shot count based on the entropy of the probability distribution, DDS effectively decreases the overall number of measurement shots without compromising fidelity. In each iteration, DDS calculates the entropy of the probability distribution and uses this information to determine the number of shots for the subsequent iteration.

\begin{figure} [t] 
  \includegraphics [width=\columnwidth] {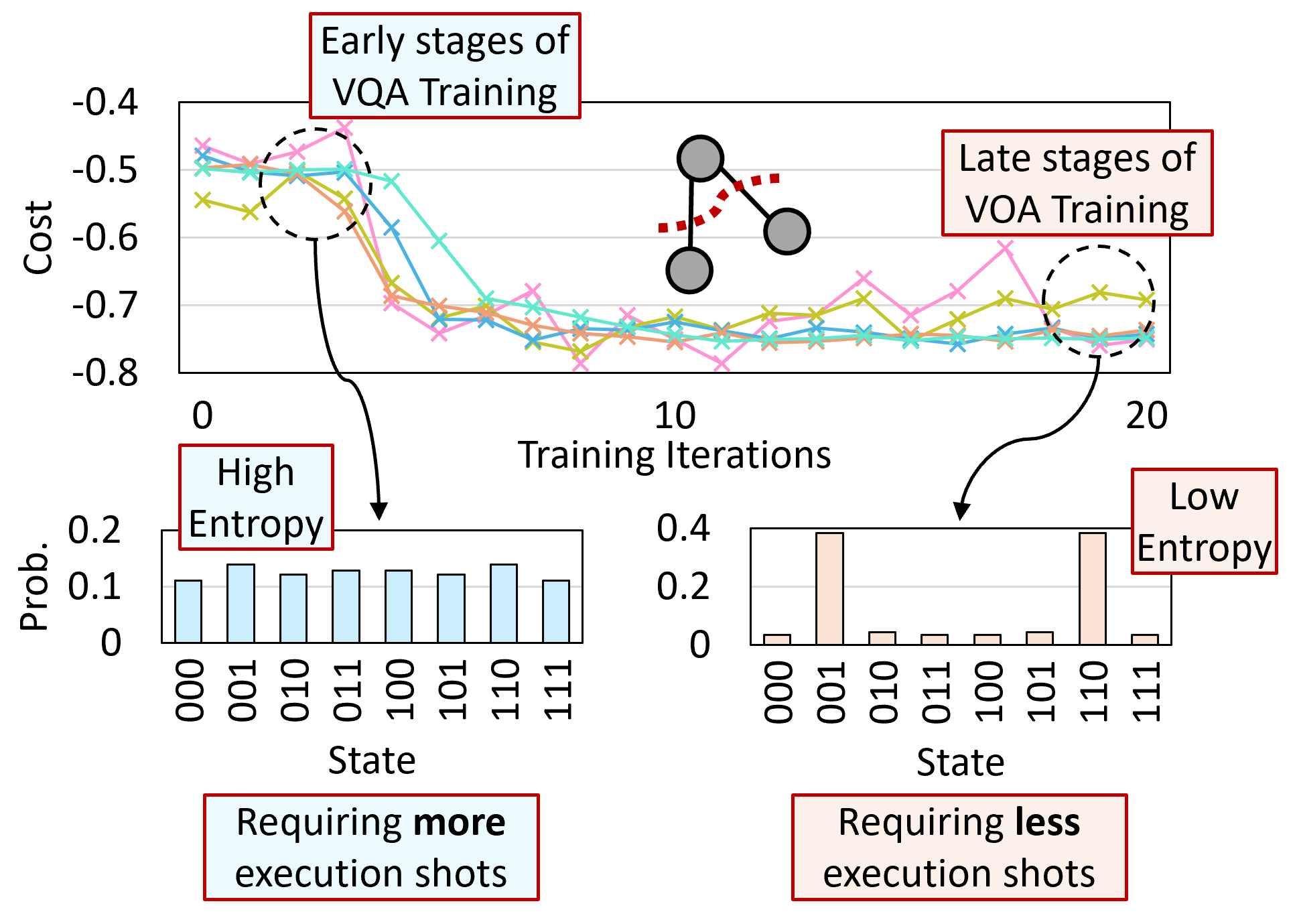}
  \caption {
   The required number of shots to obtain the probability distribution varies depending on the information entropy associated with quantum states at each iteration of the VQA.
  } 
  \label{f1}
\end{figure}

Our approach dynamically adjusts the shot count for the quantum circuit of VQAs by exploiting the entropy of probability distribution obtained from the previous iteration. Unlike existing tiered shot allocation methods that rely on fixed heuristics, the DDS scheme leverages the gradual evolution of the probability distribution throughout the VQA process. As VQA gradually refines its parameters over multiple iterations, DDS maintains simulation accuracy while enhancing implementation efficiency. By dynamically adjusting the shot count for each iteration based on the estimated probability distribution, the DDS approach reduces the overall number of shots required to optimize the variational parameters. This reduction in shot count per iteration can lead to a substantial decrease in total execution time, further enhancing the overall efficiency of VQA implementations on quantum hardware.

Fig. \ref{f1} illustrates the process of determining the number of shots associated with each iteration throughout the VQA process. We observed that as the VQA training progresses, the information entropy of the probability distribution generally decreases. During iterations with low-entropy distributions, trainability can be maintained by measuring with fewer shots while preserving accuracy. To evaluate the precision of VQA execution under adaptive shot allocation, we used the Approximation Ratio Gap (ARG) \cite{ayanzadeh2023frozenqubits}. The ARG quantifies the difference between the final cost and the ideal cost, with a lower ARG indicating higher accuracy due to closer alignment with the ideal.

The DDS approach achieves a significant reduction in the total number of shots, with an average decrease of 50.69\% compared to the baseline, even as the ARG shows a slight increase of 6.03\% on average. Although the total number of iterations remains similar, DDS notably reduces the number of shots required per iteration, contributing to an overall decrease in computational resources. When compared to the existing tiered shot allocation method, DDS requires a higher number of shots; however, it provides a substantial improvement in accuracy, achieving a 63.34\% increase in performance. Under noisy conditions, DDS achieves an average 31.15\% reduction in the total number of shots, while maintaining a slight increase in ARG compared to the baseline. Additionally, DDS provides a 68.91\% improvement in ARG compared to the tiered-shot approach, despite requiring more shots. These results highlight the effectiveness of the DDS approach in balancing shot allocation and measurement precision during VQA training.
\section{\label{sec:Observation}Entropy in Quantum Circuit Implementations}
In this section, we present our observations regarding the probability distributions and the corresponding number of shots required for quantum measurements in VQAs. Specifically, we analyze the number of shots needed to accurately capture the probability distribution of the final quantum state and the implications for the overall VQA simulation process. Additionally, we discuss the relationship between the entropy of the probability distribution and the required number of shots, as well as how this relationship affects the efficiency of VQA implementations.

\subsection{\label{sec:Observation-A}Information Entropy of Probability Distribution}
Information entropy, also referred to as \textit{Shannon entropy} \cite{shannon1948mathematical}, quantifies the uncertainty associated with a random variable. Information about events that are less likely to occur is considered more informative than information about events that happen frequently. The greater the uncertainty, the more information can be obtained, and thus entropy is defined as the mathematical expected value of the information across all possible events. Let $X$ be a random variable, which is a function of the probabilities in the sample space. For probability distributions $p_1$,..., $p_n$, the information entropy can be expressed as:
\begin{equation} \label{e1}
H(X) \equiv H(p_1, \ldots, p_n) \equiv -\sum_x p_x \log_2 p_x.
\end{equation}

The entropy of a probability distribution is high for a uniform distribution and low for distributions that are unevenly distributed. In the context of a quantum circuit, the entropy of the Walsh–Hadamard transform applied to the initial $\lvert 0 \rangle$ state is greater than the entropy of the GHZ state \cite{greenberger1989going}, which exhibits a more concentrated distribution with less uncertainty. Appendix \ref{sec:Appendix-C} presents the entropy of the random number generation circuit, which is uniformly distributed and exhibits the maximum entropy among circuits with the same number of qubits. For simplicity, we will refer to information entropy simply as entropy throughout this paper.

\subsection{\label{sec:Observation-B}Required Shots for Achieving Accurate Probability Distributions}
\begin{figure} [t] 
  \includegraphics [width=0.85\columnwidth] {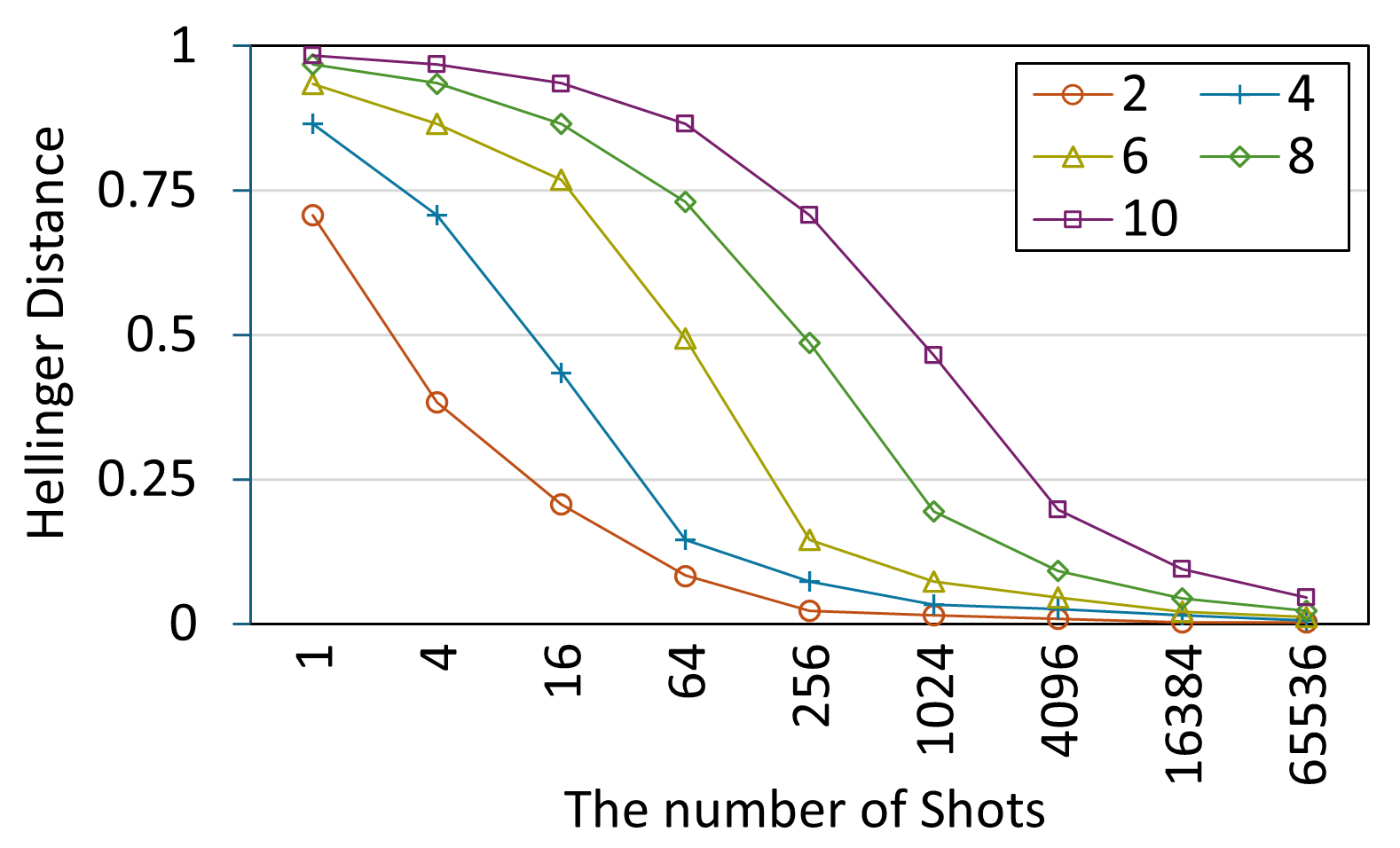}
  \caption {
   Hellinger distance of random number generation circuit \cite{li2023qasmbench} with varying qubit counts and the number of shots.
  } 
  \label{f2}
  \centerline {
  \includegraphics [width=0.85\columnwidth] {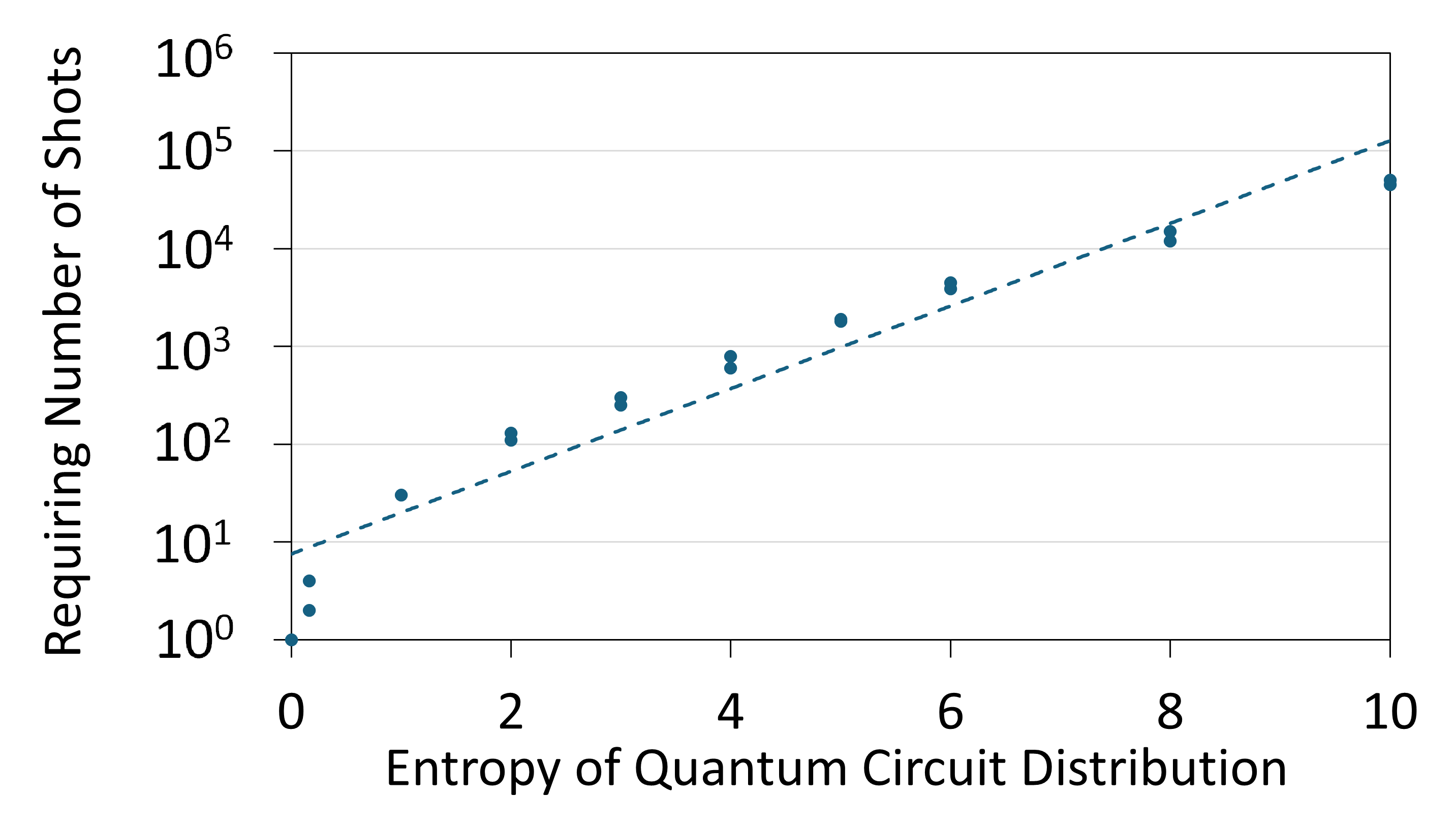} }
  \caption {
   Required number of shots with Hellinger distance not exceeding 0.05 for quantum circuit distribution entropies.
  } 
  \label{f3} 
\end{figure}

Probability distributions with more information have higher entropy; in other words, the higher the uncertainty, the higher the entropy. A uniform distribution is more informative than an unevenly distributed probability distribution that has a small number of solution states. Not only the evenness of the probability distribution but also the scalability of the sample space influences the entropy of a distribution. The sample space increases exponentially as the qubit count increases, requiring more shots to maintain accuracy. Fig. \ref{f2} shows the Hellinger distance \cite{hellinger1909neue} of the distribution generated by a random number generation quantum circuit \cite{li2023qasmbench}.

To assess the accuracy of the probability distribution of the final quantum state based on the number of measurement shots, we employ the Hellinger distance \cite{hellinger1909neue}. The Hellinger distance is a metric used to quantify the similarity between two probability distributions. A lower Hellinger distance indicates a closer match between the compared distributions, reflecting greater alignment in probability values and higher accuracy. For two probability distributions $P=(p_1,...,p_n)$ and $Q=(q_1,...,q_n)$, the Hellinger distance between $P$ and $Q$ is defined as:
\begin{equation} \label{e2}
H_d(P, Q) = \frac{1}{\sqrt{2}} \sqrt{\sum_i \left( \sqrt{p_i} - \sqrt{q_i} \right)^2}.
\end{equation}

For a fixed number of qubits, an increase in the number of measurement shots leads to a reduction in the Hellinger distance. Moreover, as the qubit count increases, the number of shots required to achieve a low Hellinger distance also increases. In circuits with higher qubit counts, a sufficiently large number of measurement shots is essential for accurately determining the quantum state. Additionally, we observe that quantum states with more informative probability distributions generally necessitate a greater number of measurement shots.

\subsection{\label{sec:Observation-C}Optimizing Shot Requirements Under a Hellinger Distance Constraint}
The outcome we want from quantum circuit execution is the probability distribution of basis states at the final state. The distribution is composed of multiple collections of basis states, which are determined for each measurement. Increasing the shot count can achieve high accuracy for any distribution, but it would take a long time to run multiple measurements. A uniform distribution represents the state with the highest entropy and the largest number of correct answers, requiring a relatively high number of measurements for accurate representation. Conversely, when the number of correct answers is smaller, the distribution can be guaranteed with a comparable level of accuracy using fewer measurements.

To maintain the Hellinger distance while reducing the number of shots, it is essential to determine the minimum number of shots required for each probability distribution. Fig. \ref{f3} illustrates the relationship between entropy and the number of shots needed to approximate the ideal probability distribution that the quantum state is theoretically expected to output. The x-axis represents the entropy of the output distribution, while the y-axis shows the number of shots required to achieve a Hellinger distance of 0.05. The results reveal an exponential relationship between the entropy of the target distribution and the number of shots necessary to achieve a specified level of accuracy. Therefore, by referencing the entropy of a distribution in VQA iterations, we can estimate the required shot count to ensure accurate measurement.
\section{\label{sec:DDS}Distribution-Adaptive Dynamic Shot (DDS)}
In this section, we introduce an optimization method based on the distribution of quantum states that dynamically adjusts the shot count per iteration in VQAs to reduce the total number of measurements. We will discuss previous observations in the context of VQAs and show how these insights are leveraged in the DDS technique. Subsequently, we will provide a detailed explanation of the DDS approach.

\subsection{\label{sec:DDS-A}Evolution of Probability Distributions in VQA}

In VQA, the quantum circuit execution and classical optimization of circuit parameters are performed iteratively. For each iteration, measurements are taken based on the specified number of shots, with the resulting probability distribution derived from the measurement outcomes. Each measurement corresponds to a computational basis state, and the proportion of results that align with a specific basis relative to the total number of measurements, defines the probability of that basis. As the quantum circuit parameters are updated, the probability distribution dynamically evolves with each iteration. Therefore, the probability distribution changes throughout the VQA process, with these variations being influenced by the algorithms or ansatz employed. A more detailed overview of VQA can be found in Appendix \ref{sec:Appendix-B}.

For most VQA training or other quantum algorithms, the initial circuit preparation is typically performed by applying a Hadamard gate to the $\lvert 0 \rangle$ state multiple times. At the beginning of the VQA training iterations shown in Fig. \ref{f1}, the probability distributions are likely to be evenly distributed. For the QAOA max-cut problem, the final state contains information about the edges for the max cut, and the number of correct states varies depending on the QAOA graph. In cases where edges are densely connected, the probability distribution will be evenly distributed, whereas in cases where edges are concentrated on one vertex or in a graph with low density, the probability distribution will be concentrated on a few correct states. At the beginning of the training iterations, a large number of shots will be required, while in later training iterations, when the distribution becomes unevenly distributed, only a small number of shots will be needed.

\subsection{\label{sec:DDS-B}Mechanics of Distribution-Adaptive Shot Allocation}

In VQA iterations, DDS evaluates the probability distributions and dynamically adjusts the number of measurement shots in each iteration to effectively minimize the total shot count. The probability distribution at each iteration is incrementally modified by the classical optimizer, reflecting changes in the quantum circuit parameters. Since consecutive iterations typically exhibit similar distributions, DDS leverages the entropy of the probability distribution from the previous iteration to determine the required shot count for the current iteration, ensuring accurate and reliable operation of the VQA process.

For each iteration, measurements are performed based on the specified shot count, and the resulting probability distribution is derived from the measurement outcomes. These outcomes, corresponding to measurement bases (commonly computational bases), are stored as counts. The probability distribution is then constructed from these counts, enabling the entropy $H$ to be calculated in the estimator. This entropy serves as a key metric for effectively estimating the shot count required in subsequent iterations. For each measurement basis $A_j$, the entropy of the probability distribution is defined as:
\begin{equation} \label{e3}
H = -\sum_j P(A_j) \log_2 P(A_j).
\end{equation}

The entropy $H$ is stored and utilized in the subsequent iteration to estimate the shot count. After the estimation process for the following iteration is completed, the calculated entropy is updated accordingly. This entropy calculation is repeated for each iteration until the parameters converge, as determined by the classical optimizer.

Determining the shot count for initializing the estimator is a crucial component of the DDS method. If no specific shot count is provided, the estimator defaults to 1,024 shots. DDS dynamically adjusts the number of shots for each iteration using the entropy value stored from the previous iteration. Based on the observation that the required number of shots is exponentially proportional to the entropy, as discussed in Section-\ref{sec:Observation}, the shot count $S$ is calculated from the entropy $H$ as follows:
\begin{equation} \label{e4}
S = k \times 2^{H},
\end{equation}
where $k$ is a constant. The value of $k$ depends on the type of algorithm or ansatz used. The number of qubits also influences the selection of $k$: large circuits typically require more shots, as noted in Section-\ref{sec:Observation}. For circuits with many qubits, where a significant number of shots is needed to achieve a sufficiently low Hellinger distance, $k$ should be set to a larger value. Conversely, for circuits that require fewer shots, setting $k$ to a smaller value is more appropriate to minimize the overall shot count.

Algorithm \ref{alg:VQA_DynamicShot} outlines the process of dynamically adjusting the number of shots during the VQA process. At each iteration, the entropy of the probability distribution obtained from the measurement results in the previous iteration is calculated. This entropy value is then used to determine the number of shots required for the current iteration. By dynamically adapting the shot count based on the entropy, the algorithm effectively reduces the number of shots per iteration while maintaining the accuracy of the resulting probability distribution.

\begin{algorithm}[t]
\caption{VQA with Distribution-Adaptive Dynamic Shot Allocation}
\label{alg:VQA_DynamicShot}
\KwIn{ansatz, initial\_parameter}
\KwOut{\emph{optimal\_parameter}}

\medskip
\DontPrintSemicolon

parameter $\gets$ initial\_parameter\;
prev\_entropy $\gets$ 10\ \tcp{initialize entropy};
converged $\gets$ False\;
\While{not converged}{
    entropy $\gets$ prev\_entropy\;
    shots $\gets$ constant $\times$ $2^{\text{entropy}}$\;   
    result $\gets$ Estimator(ansatz, parameter, shots)\;
    prev\_entropy $\gets$ ComputeEntropy(result)\;
    cost $\gets$ expectation\_value(result)\;    
    parameter, converged $\gets$ cobyla(ansatz, result)\;   
}
\end{algorithm}
\section{\label{sec:Evaluation}Evaluation}

\subsection{\label{sec:Evaluation-A}Evaluation Methodology}

In this section, we compare the proposed DDS method with the standard fixed-shot approach. Furthermore, we evaluate its performance relative to existing shot reduction strategies: the linear function-based and step function-based measurement shot allocation methods.

\subsubsection{\label{sec:Evaluation-A-1}Scenarios}

The fixed-shot method uses 1,024 shots as its configuration, aligning with the default setting in many existing quantum computing systems. The linear shot function \cite{phalak2023shot}, starting from 1,000 shots (close to the default of 1,024), gradually reduces the shot count over training iterations. This function is defined as $S_{i} = \max(20, S_{begin} - l \cdot i)$, where $S_{i}$ represents the shot count at the $i^{th}$ iteration, and $l$ is the slope parameter that controls the rate of reduction. In this study, we set $S_{begin} = 1,000$ and $l = 10$. To prevent impractically low shot counts, we enforce a minimum threshold of 20 shots. Similarly, the step function \cite{phalak2023shot} reduces the shot count in discrete increments and is defined as $S_{i} = \max(20, S_{begin} - 10 \cdot l \cdot \left\lfloor \frac{i}{10} \right\rfloor)$, where $\left\lfloor \cdot \right\rfloor$ denotes the floor function. To ensure consistency with the linear shot function, we use $S_{begin} = 1,000$ and set the slope parameter to $l = 10$. These values for $S_{begin}$ and the slope parameter were applied uniformly across all VQA applications during the evaluation.

\subsubsection{\label{sec:Evaluation-A-2}Benchmarks and Backends}

To evaluate the performance of the DDS approach, quantum circuits for both the Quantum Approximate Optimization Algorithm (QAOA) and the Variational Quantum Eigensolver (VQE) were constructed. For QAOA, we selected target graphs commonly used in network science, specifically Power-Law (PL), Barabási–Albert (BA), Watts–Strogatz (WS), and Sherrington–Kirkpatrick (SK) models. Each graph model exhibits a unique connection density between nodes. For all target graphs, the QAOA circuit was configured with 20 layers, and models with varying qubit counts were tested.

For the VQE experiments, we simulated the ground state energies of H$_{2}$, LiH, and BeH$_{2}$ in the STO-3G basis set. The VQE circuits were configured to meet the qubit requirements for each molecule. We used the Unitary Coupled Cluster with Single and Double excitations (UCCSD) ansatz for VQE simulations, which requires 4, 12, and 14 qubits for H$_{2}$, LiH, and BeH$_{2}$, respectively. This approach allows us to map spin orbitals to individual qubits, and we use all orbitals without freezing the core orbitals. The VQE calculations using the UCCSD ansatz were performed with bond lengths of 0.73{\AA} for H$_{2}$, 1.57{\AA} for LiH, and 3.00{\AA} for BeH$_{2}$.

All VQA benchmark circuits were simulated on the qiskit-aer statevector simulator \cite{ibm}, where dynamic shot allocation methods, including DDS, implemented with few modifications. For experiments conducted under noisy conditions, the qiskit-aer qasm simulator was employed. This noise simulation environments reflect the 1Q and 2Q gate error rates of ibm\_marrakesh with Heron r2 and ibm\_yonsei with Eagle r3, calibrated on December 13, 2024. In both QAOA and VQE experiments, the classical optimizer COBYLA was applied, given its widespread application in variational quantum algorithms.

\subsubsection{\label{sec:Evaluation-A-3}Metrics}

To measure the fidelity of VQA circuits, Approximation Ratio Gap (ARG) 
 \cite{ayanzadeh2023frozenqubits} was used, defined as follows:
\begin{equation} \label{e5}
ARG = 100 \times \left\vert\frac{E_{ideal}-E_{real}}{E_{ideal}}\right\vert.
\end{equation}
$E_{ideal}$ is the ideal expectation value of the quantum state, and $E_{real}$ is the calculated expectation value. When the fidelity is high, the measured expectation value is closer to the ideal value, resulting in a smaller ARG value, which indicates a more accurate approximation. The smaller the ARG value, the better the performance.

$S_{tot}$ is the total number of measurements in the VQA process, and $S_{avg}$ is the average number of measurements taken per iteration. For the total number of iterations $I_{tot}$, the total number of shots is calculated as follows:
\begin{equation} \label{e6}
S_{tot} = S_{avg} \times I_{tot}.
\end{equation}

\subsection{\label{sec:Evaluation-B}Results and Analysis}

\begin{figure}[b]
  \centering 
  \includegraphics [width=0.98\columnwidth] {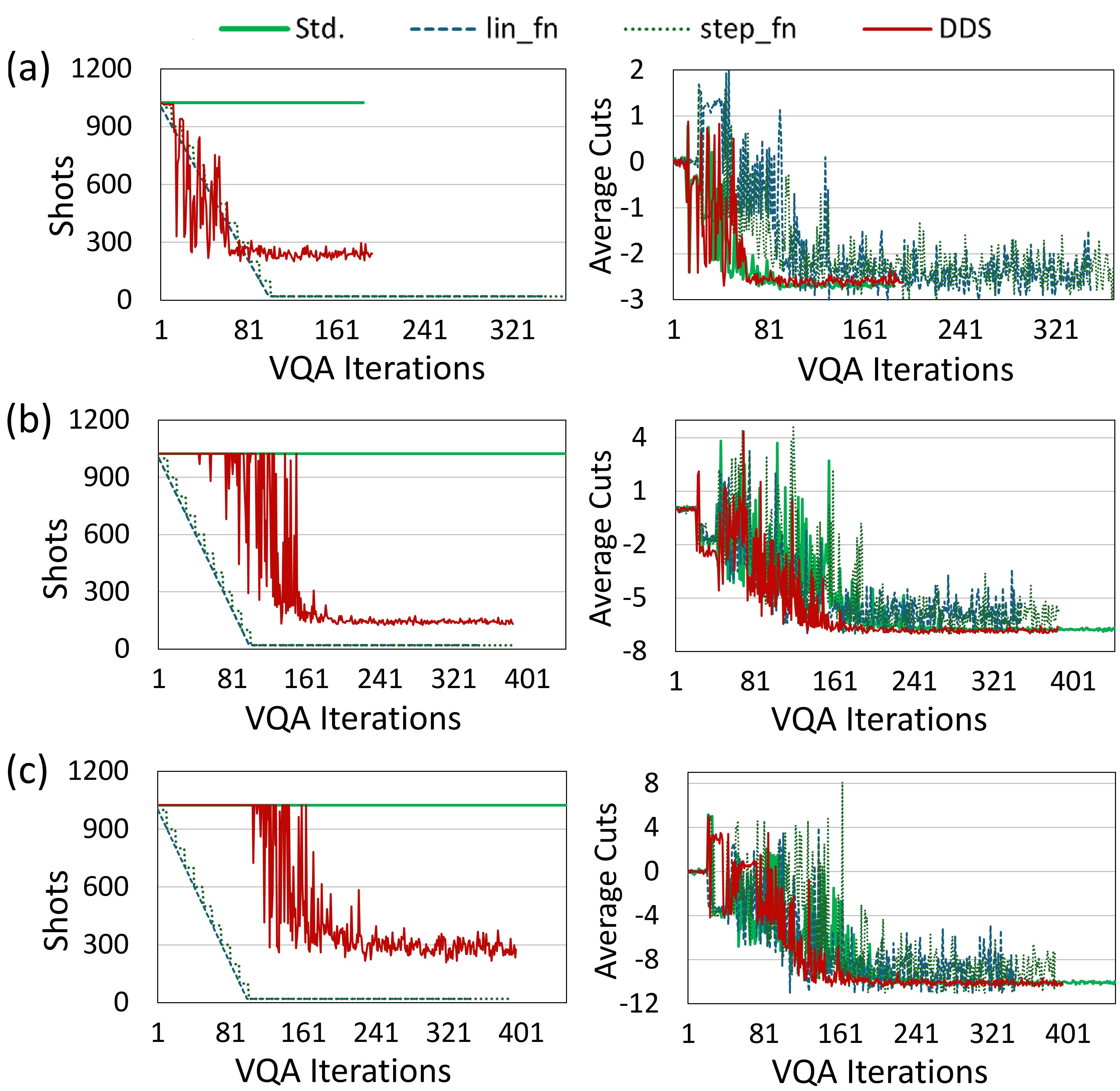}
  \caption{The number of shots and average cuts per iteration during the PL model QAOA training for (a) 4 qubits, (b) 8 qubits, and (c) 12 qubits, using different shot allocation methods.}
  \label{fig:4}
\end{figure}

\subsubsection{\label{sec:Evaluation-B-1}Evaluation on Scalable VQAs}

Dynamic shot allocation was applied to both QAOA and VQE models for several molecules. We tested four QAOA models and three VQE models, running each with the standard fixed shot method of 1,024 shots, as well as with two existing shot reduction methods: the linear function and step function-based shot allocation approaches \cite{phalak2023shot}. The standard method, a conventional approach, is labeled as ``Std." in the graphs and tables. As described earlier, the linear and step functions begin with 1,000 shots and gradually decrease the shot count over training. These methods are indicated as ``lin\_fn" and ``step\_fn" in the tables and graphs, respectively. For DDS, the implementation is the same as the approach outlined in the previous section, and an upper bound of 1,024 shots was applied to allow for direct comparison with the standard method throughout the experiments.

\begin{figure}[t]
  \centering
  \includegraphics [width=\columnwidth] {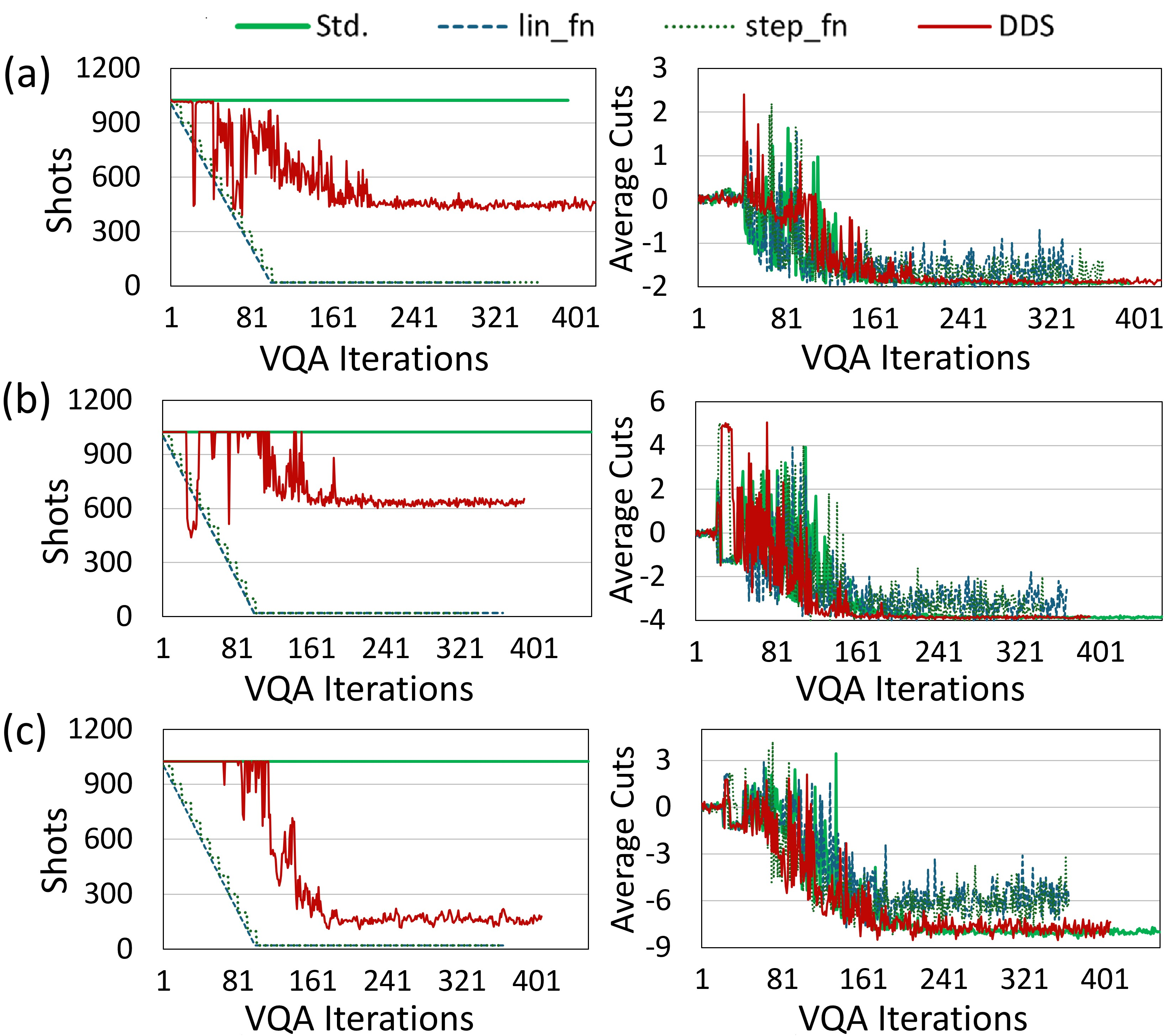}
  \caption{The number of shots and average cuts per iteration during the BA model QAOA training for (a) 4 qubits, (b) 8 qubits, and (c) 12 qubits, using different shot allocation methods.}
  \label{fig:5}
\end{figure}

\begin{figure}[t]
  \centering
  \includegraphics [width=\columnwidth] {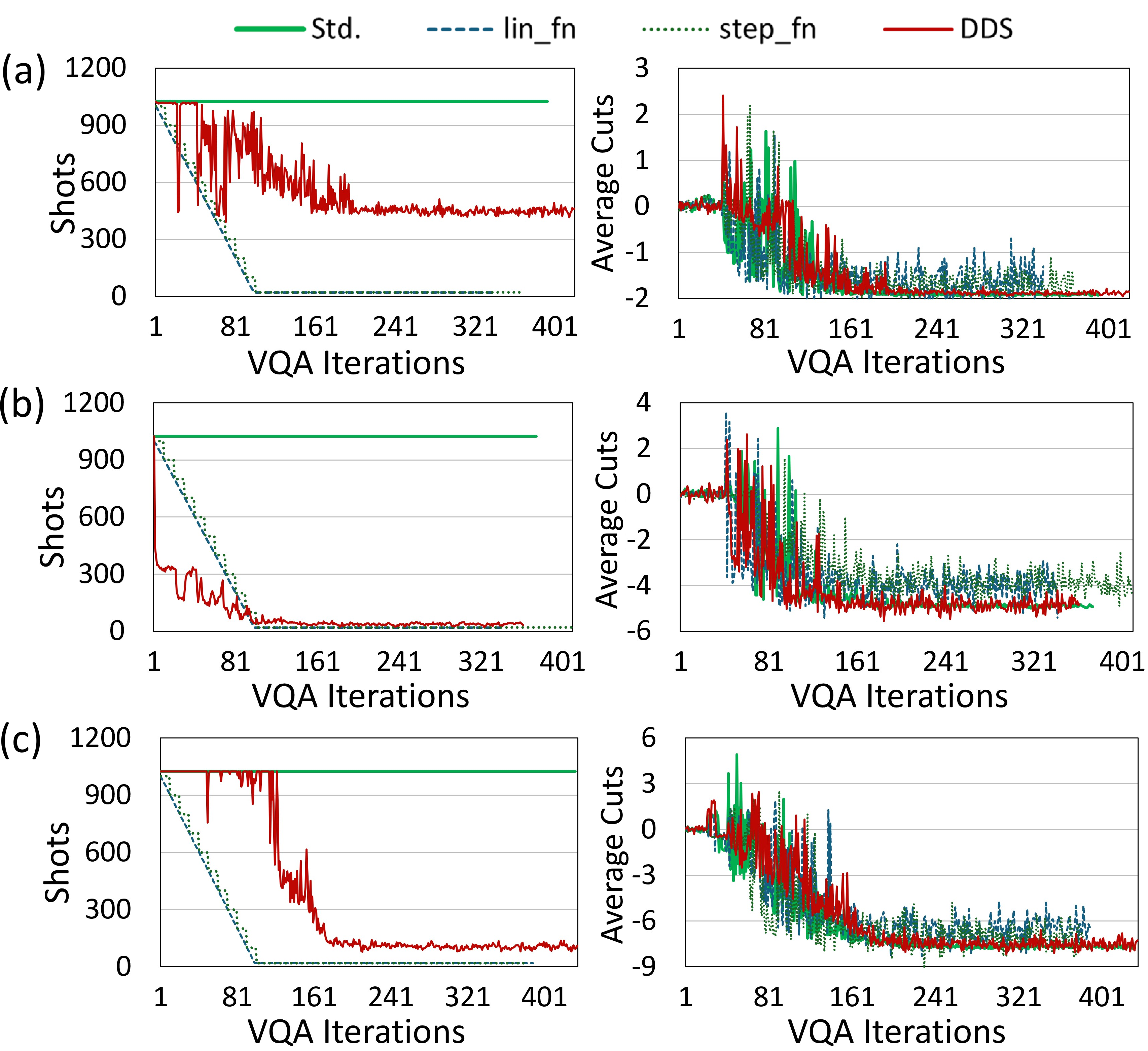}
  \caption{The number of shots and average cuts per iteration during the WS model QAOA training for (a) 4 qubits, (b) 8 qubits, and (c) 12 qubits, using different shot allocation methods.}
  \label{fig:6}
\end{figure}

Figure \ref{fig:4} illustrates the number of shots and the average cuts (i.e., the expectation value) per iteration for QAOA PL models with 4, 8, and 12 qubits, using four shot allocation methods: standard, linear\_fn, step\_fn, and DDS. Both linear\_fn and step\_fn follow a consistent downward trend until they reach the minimum shot limit of 20, maintaining that number until the end. In contrast, DDS dynamically allocates shots based on the probability distribution of the results from the previous iteration. As a result, DDS initially requires a high number of shots, which gradually decreases over the course of iterations.

\begin{figure}[t]
  \centering
  \includegraphics [width=\columnwidth] {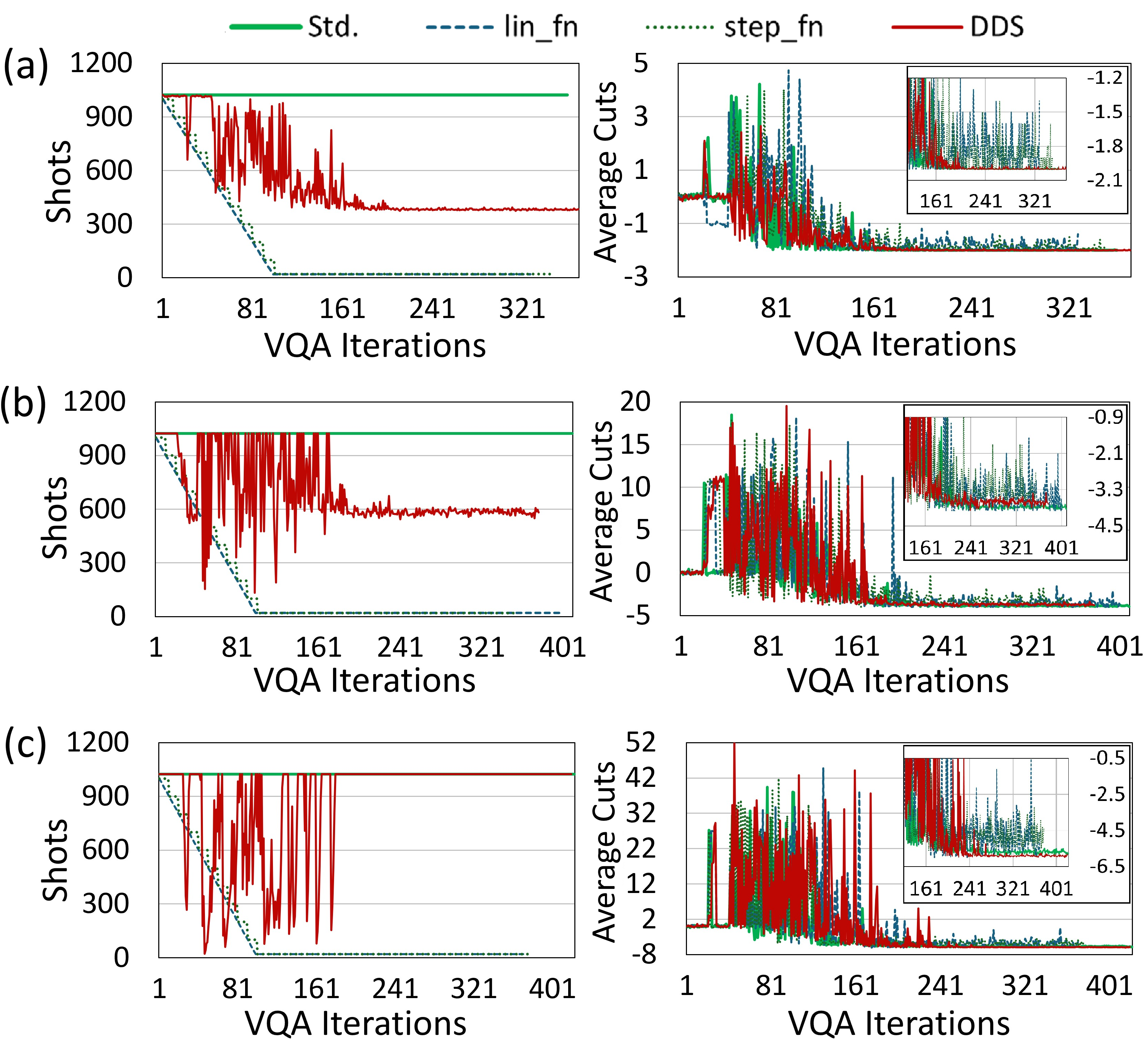}
  \caption{The number of shots and average cuts per iteration during the SK model QAOA training for (a) 4 qubits, (b) 8 qubits, and (c) 12 qubits, using different shot allocation methods.}
  \label{fig:7}
\end{figure}

As the QAOA process advances, the quantum state approaches the solution, becoming more concentrated on certain states. Consequently, entropy decreases, leading to a reduced shot requirement in later iterations. When examining the cost (expectation value) across iterations, DDS performs comparably to the standard method. However, for linear\_fn and step\_fn, the cost does not converge consistently, displaying oscillations. While the total shot count may be lower for linear\_fn and step\_fn, the reduced number of shots hinders their ability to reliably converge to a low expectation value.

Figures \ref{fig:5}, \ref{fig:6}, and \ref{fig:7} display the shot count and the average cuts per iteration for the QAOA BA, WS, and SK models, respectively, under each shot allocation method. Similar to the PL model in Figure \ref{fig:4}, DDS begins with a high shot count and decreases it in later iterations. In terms of cost, DDS remains close to the performance of the standard method, whereas linear\_fn and step\_fn show fluctuations and fail to converge effectively to low expectation values.

For the SK model, due to its higher graph complexity, the required shot count per iteration is greater than for other models, and the cost also exhibits more significant fluctuations. In linear\_fn and step\_fn, cost fluctuations are more pronounced initially, but for models with lower graph complexity, DDS demonstrates less variability in cost. Given that the primary objective in QAOA is to achieve low expectation values, linear\_fn and step\_fn may face difficulties in achieving stable and consistent performance. In contrast, DDS effectively produces results comparable to the standard method in a stable manner.

Table \ref{Table:t1} in Appendix \ref{sec:Appendix-D} presents the average shot count ($S_{avg}$) for each QAOA model based on qubit count, the total number of iterations ($Iterations$) over which QAOA is simulated, and the $Average$ $Cuts$, which represents the result of an additional execution following the last iteration. Due to variations in the shot allocation method, the number of shots used in the final iteration may differ. To ensure a fair comparison, this final execution was conducted with a fixed shot count of 1,024 for all methods.

\begin{figure}[t]
  \centering
  \includegraphics [width=\columnwidth] {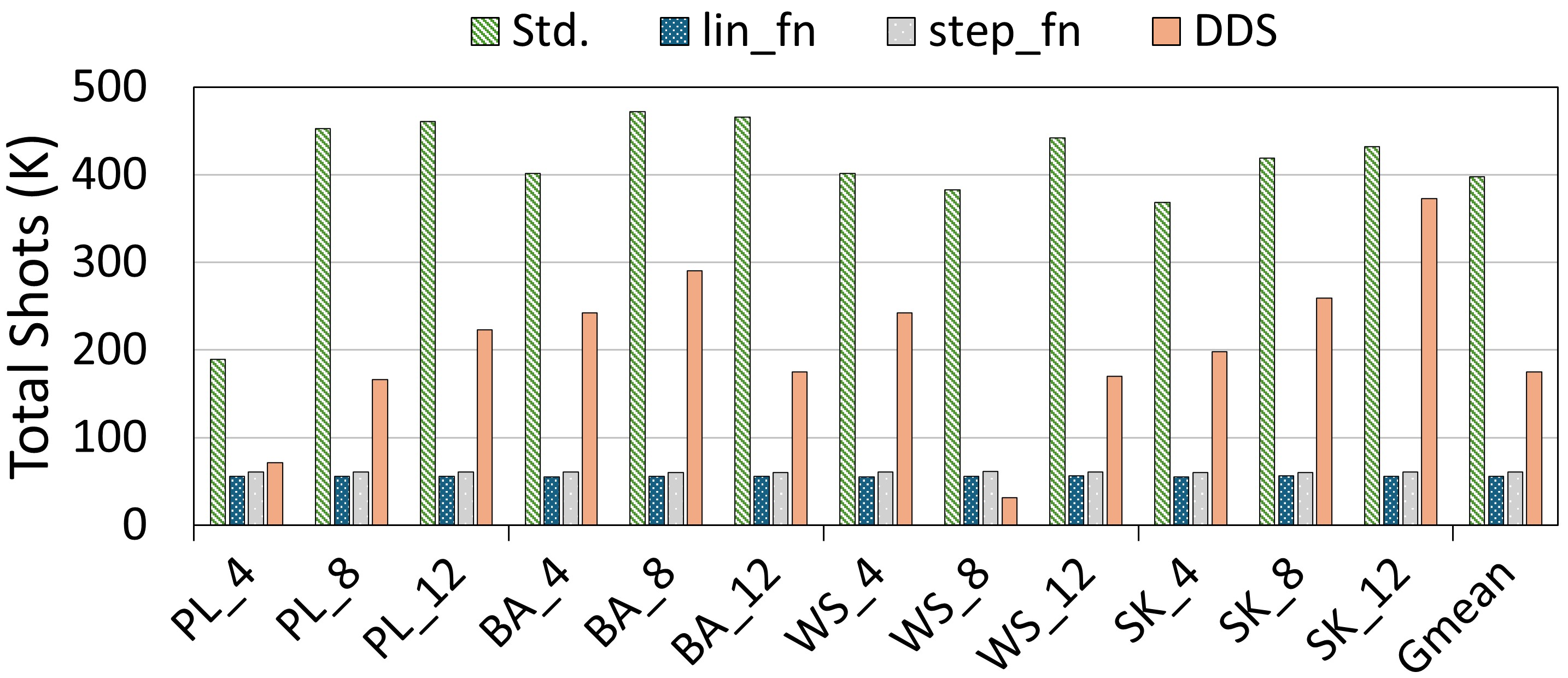}
  \caption{Total shot counts required for the training of various QAOA models, including Power-Law (PL), Barabási–Albert (BA), Watts–Strogatz (WS), and Sherrington–Kirkpatrick (SK), across different qubit counts and model configurations.}
  \label{fig:8}
\end{figure}

\begin{figure}[t]
  \centering
  \includegraphics [width=\columnwidth] {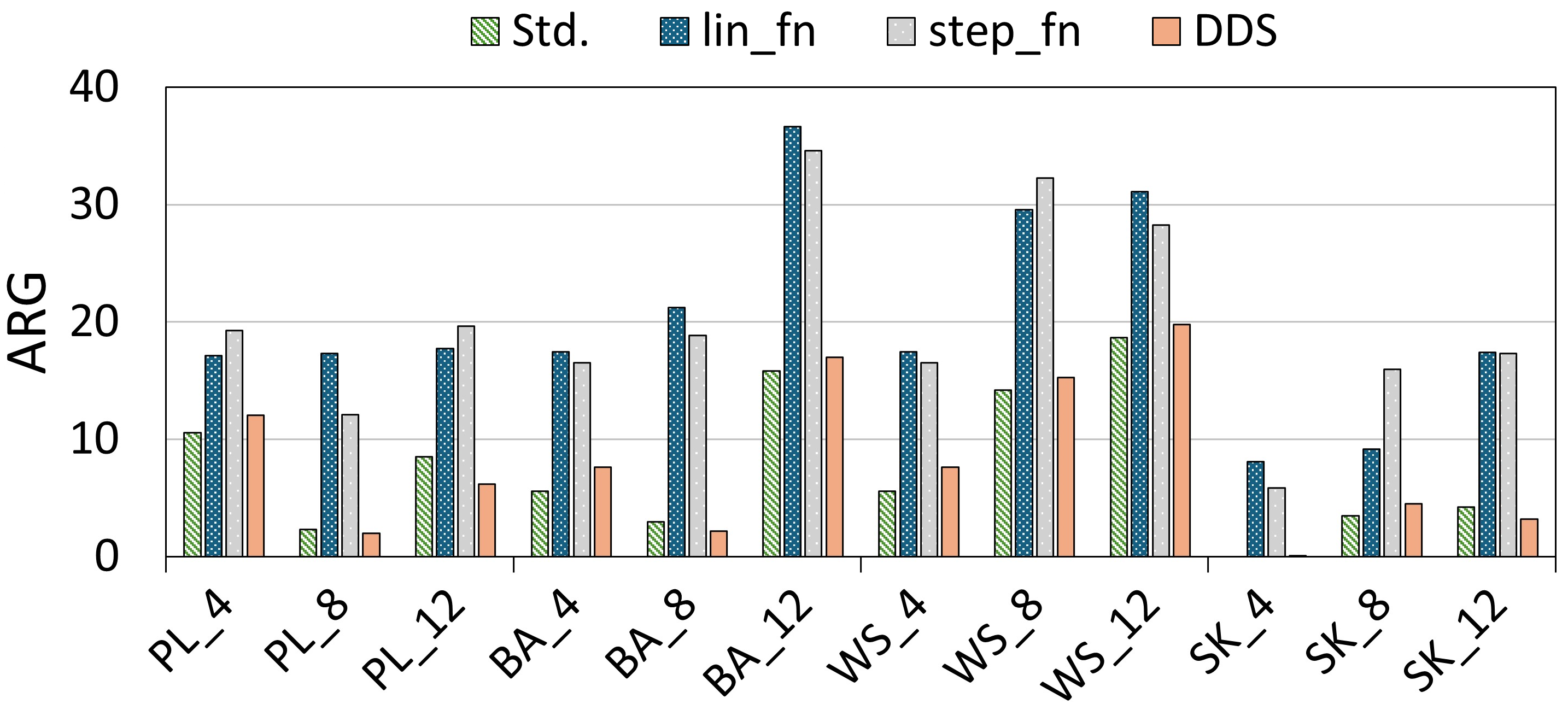}
  \caption{The ARG values for the training of various QAOA models, including Power-Law (PL), Barabási–Albert (BA), Watts–Strogatz (WS), and Sherrington–Kirkpatrick (SK), across different qubit counts and model configurations.}
  \label{fig:9}
\end{figure}

For QAOA, the total number of iterations shows no substantial difference across methods. However, the average shot count per iteration is noticeably lower for DDS compared to the standard method. Figure \ref{fig:8} illustrates the total number of shots for each shot allocation method, indicating that DDS, on average, requires 50.69\% fewer shots than the standard method. Meanwhile, the linear\_fn and step\_fn methods achieve greater reductions, with the total shot count decreasing by 85.49\% and 84.24\%, respectively, compared to the standard method.

To assess the fidelity of QAOA, the ARG metric was examined. As shown in Figure \ref{fig:9}, the ARG values for linear\_fn and step\_fn are substantially higher than the standard, with average increases of 241.18\% and 230.73\%, respectively. In contrast, the ARG for DDS remains close to the standard, showing only a 6.03\% average difference. Thus, DDS achieves a substantial reduction in shot count with minimal fidelity loss. In contrast, although the linear\_fn and step\_fn methods reduce shot counts considerably, they incur significant fidelity losses, which may limit their suitability for accurate QAOA simulation. Compared to these tiered shot allocation methods, DDS demonstrates a 63.34\% improvement in accuracy.

\begin{figure}[t]
  \centering
  \includegraphics [width=\columnwidth] {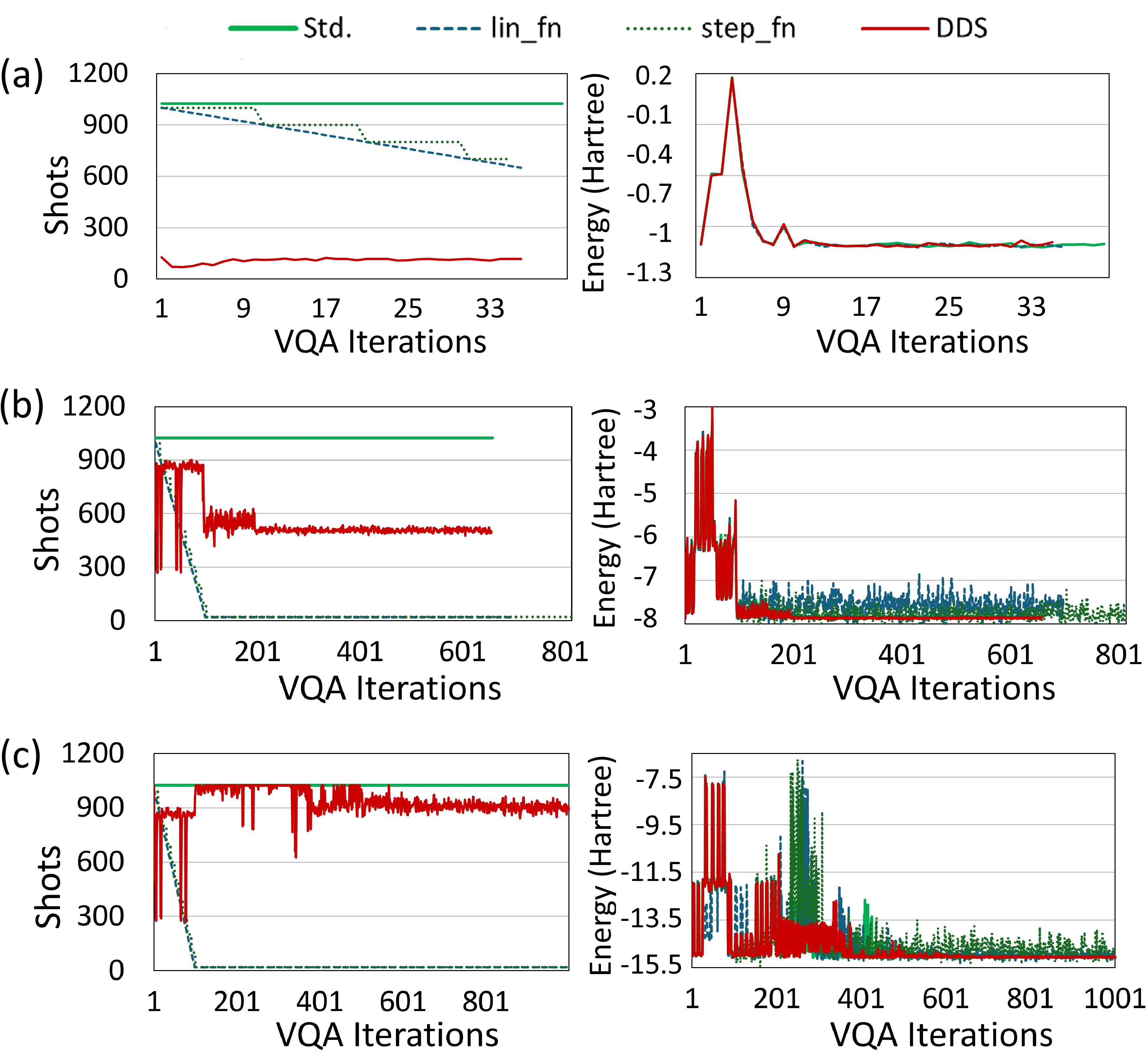}
  \caption{The number of shots and the energy of each iteration for the VQE training process of (a) H$_{2}$, (b) LiH, and (c) BeH$_{2}$.}
  \label{fig:10}
\end{figure}

Figure \ref{fig:10} illustrates the shot count and the energy per iteration for VQE simulations on H$_{2}$, LiH, and BeH$_{2}$ molecules, using the standard, linear\_fn, step\_fn, and DDS shot allocation methods. In these simulations, the expectation value computed through the VQE circuit represents the electronic ground state energy, with the total ground state energy obtained by adding the nuclear repulsion energy. For all three molecular structures, linear\_fn and step\_fn exhibit a steady initial reduction in shot count. However, as molecular complexity and qubit requirements increase, more iterations are needed for convergence. Consequently, when the shot count reaches the minimum too early, the cost fails to converge and instead fluctuates. In the case of H$_{2}$, the entropy of the distribution is low, allowing DDS to require fewer shots while maintaining a cost almost identical to the standard. For both LiH and BeH$_{2}$, the DDS method maintains a cost trajectory similar to the standard with minimal oscillations, demonstrating stable convergence.

Table \ref{Table:t2} in Appendix \ref{sec:Appendix-D} summarizes the average shot count per iteration, the total number of iterations required for the VQE process, and the final total ground state energy after convergence. To ensure consistent comparison, the energy was calculated after one additional execution with 1,024 shots across all shot allocation methods, same as the QAOA setup. For the VQE, the total number of iterations shows minimal difference across methods. However, the average number of shots per iteration varies significantly. As the complexity of the molecular ansatz increases, so does the number of required iterations. Consequently, while the average shot count per iteration for linear\_fn and step\_fn decreases over time, the average shot count for DDS increases as the entropy of the probability distribution grows, reflecting the increased uncertainty in the quantum state throughout the optimization process.

\begin{figure}[t]
  \centering
  \includegraphics [width=\columnwidth] {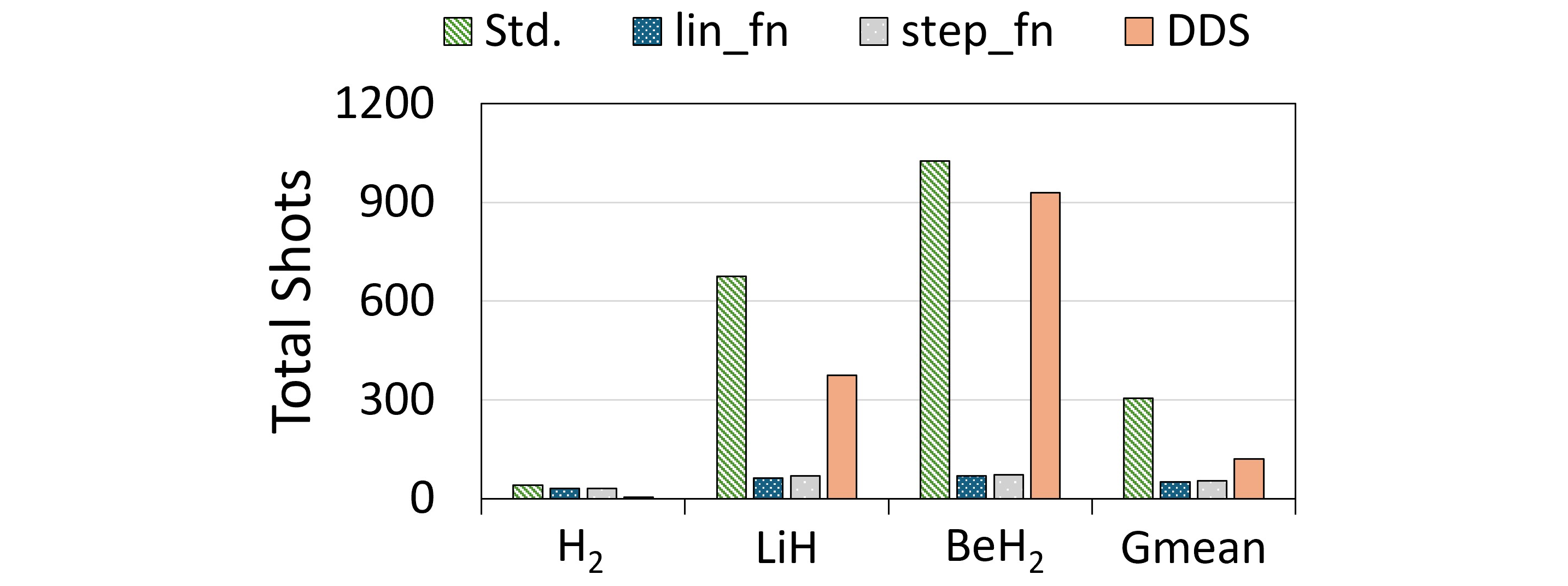}
  \caption{Total shot counts required for the training of various molecules in VQE simulations, including H$_{2}$, LiH, and BeH$_{2}$.}
  \label{fig:11}
\end{figure}

\begin{figure}[t]
  \centering
  \includegraphics [width=\columnwidth] {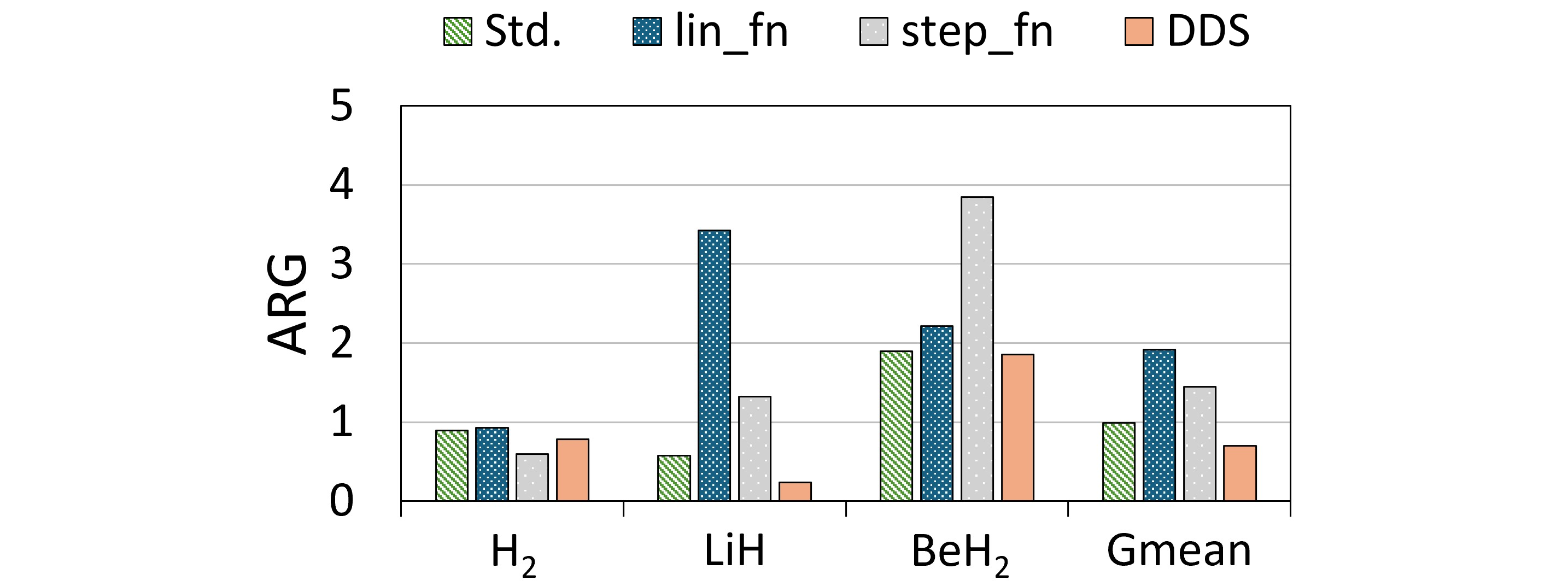}
  \caption{The ARG values for the training of various molecules in VQE simulations, including H$_{2}$, LiH, and BeH$_{2}$.}
  \label{fig:12}
\end{figure}

Figure \ref{fig:11} compares the total number of shots across shot allocation methods for the VQE process. As molecular complexity increases, reflected in the size and number of atoms in the molecular ansatz, linear\_fn and step\_fn exhibit relatively small changes in total shot count compared to other methods, with average reductions of 69.81\% and 68.64\%, respectively, compared to the standard. For LiH and BeH$_{2}$, shot reductions approach nearly 90\%. DDS achieves an average reduction of 47.09\%, with H$_{2}$ and LiH showing reductions of nearly 65\%, though it still requires more shots than linear\_fn and step\_fn due to its dynamic convergence behavior.

In terms of ARG, as illustrated in Figure \ref{fig:12}, linear\_fn and step\_fn show increases of 172.65\% and 66.79\% compared to the standard, respectively. In contrast, DDS achieves a 24.39\% reduction in ARG. For VQE, while linear\_fn and step\_fn significantly reduce shot count, they experience a notable increase in ARG, indicating reduced fidelity. DDS, however, not only decreases the number of shots but also maintains ARG levels comparable to the standard, demonstrating its effectiveness in preserving fidelity while reducing shot count. Compared to the tiered shot allocation methods, DDS demonstrates a 37.78\% improvement in accuracy. The results indicate that linear\_fn and step\_fn reduce shot counts but at the cost of significantly reduced fidelity. Thus, DDS is a more reliable approach for achieving precise quantum solutions.

\subsubsection{\label{sec:Evaluation-B-2}Evaluation under Noise Condition}

On actual quantum devices, experiments are typically conducted with a default configuration of 1,024 shots \cite{ibm}, similar to quantum simulators. However, due to the presence of noise in real quantum hardware, the number of shots is sometimes increased to ensure more reliable results \cite{miki2022quantum}. To simulate experiments under noisy conditions, the qiskit-aer qasm simulator was utilized, incorporating the native gate set and error information from the ibm\_marrakesh and ibm\_yonsei processors, which are based on the Heron and Eagle architectures. The noise model in the qasm simulator was configured to introduce depolarizing errors, accounting for both 1-qubit and 2-qubit gate errors.

The standard fixed-shot method and tiered shot allocation methods were used as before. Previously, the maximum shot count for DDS was set to 1,024 for comparison with other shot allocation methods. However, since a larger number of shots may be required to ensure reasonable accuracy in noisy environments, DDS\_M, with a maximum shot count of 100,000, was also included for comparison.

\begin{figure}[t]
  \centering
  \includegraphics [width=\columnwidth] {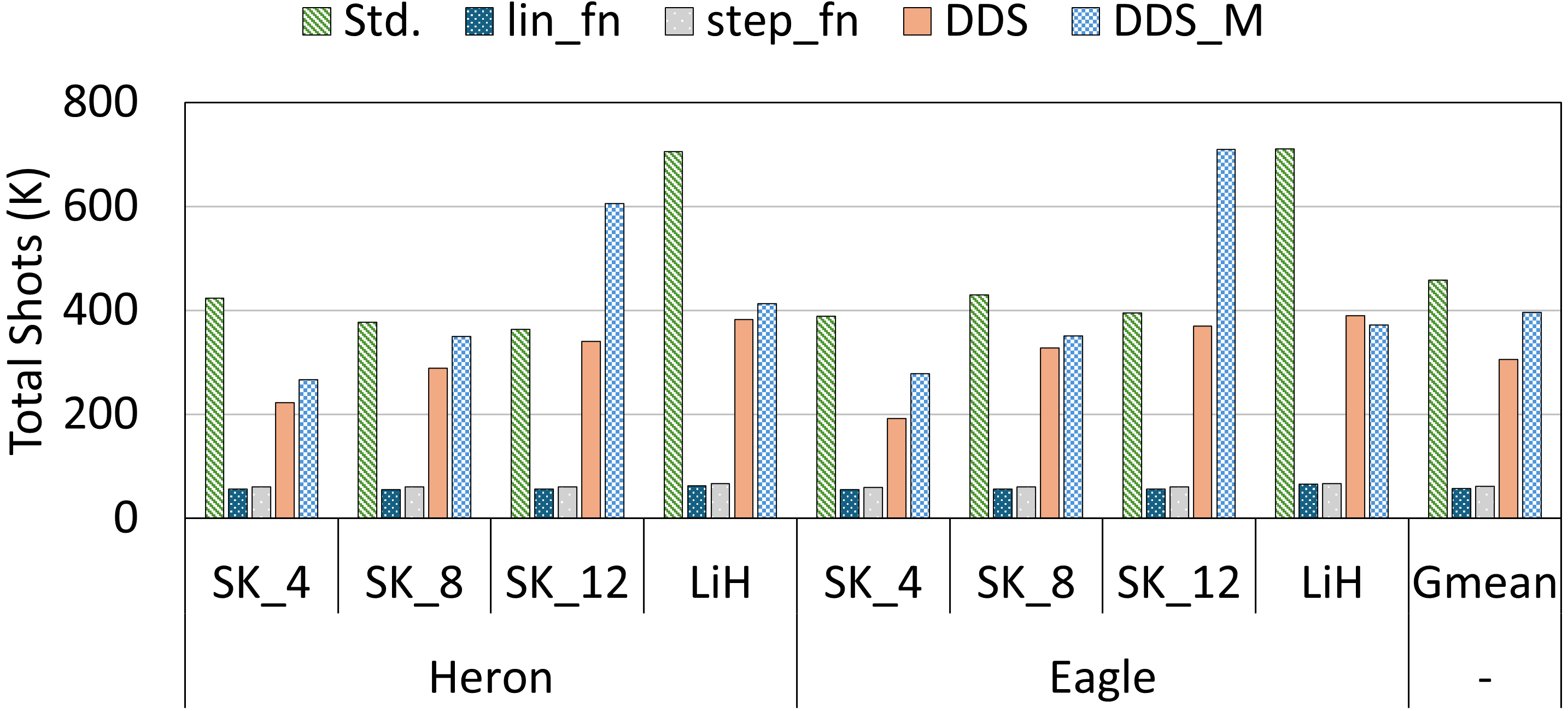}
  \caption{Total shot counts required for the training of Sherrington–Kirkpatrick (SK) QAOA model across different qubit counts and LiH VQE simulations under noisy condition. The qiskit-aer qasm simulator was used, and the native gate set along with error information from the $ibm\_marrakesh$ and $ibm\_yonsei$ processor, based on the Heron and Eagle architecture, were employed.}
  \label{fig:13}
\end{figure}

\begin{figure}[t]
  \centering
  \includegraphics [width=\columnwidth] {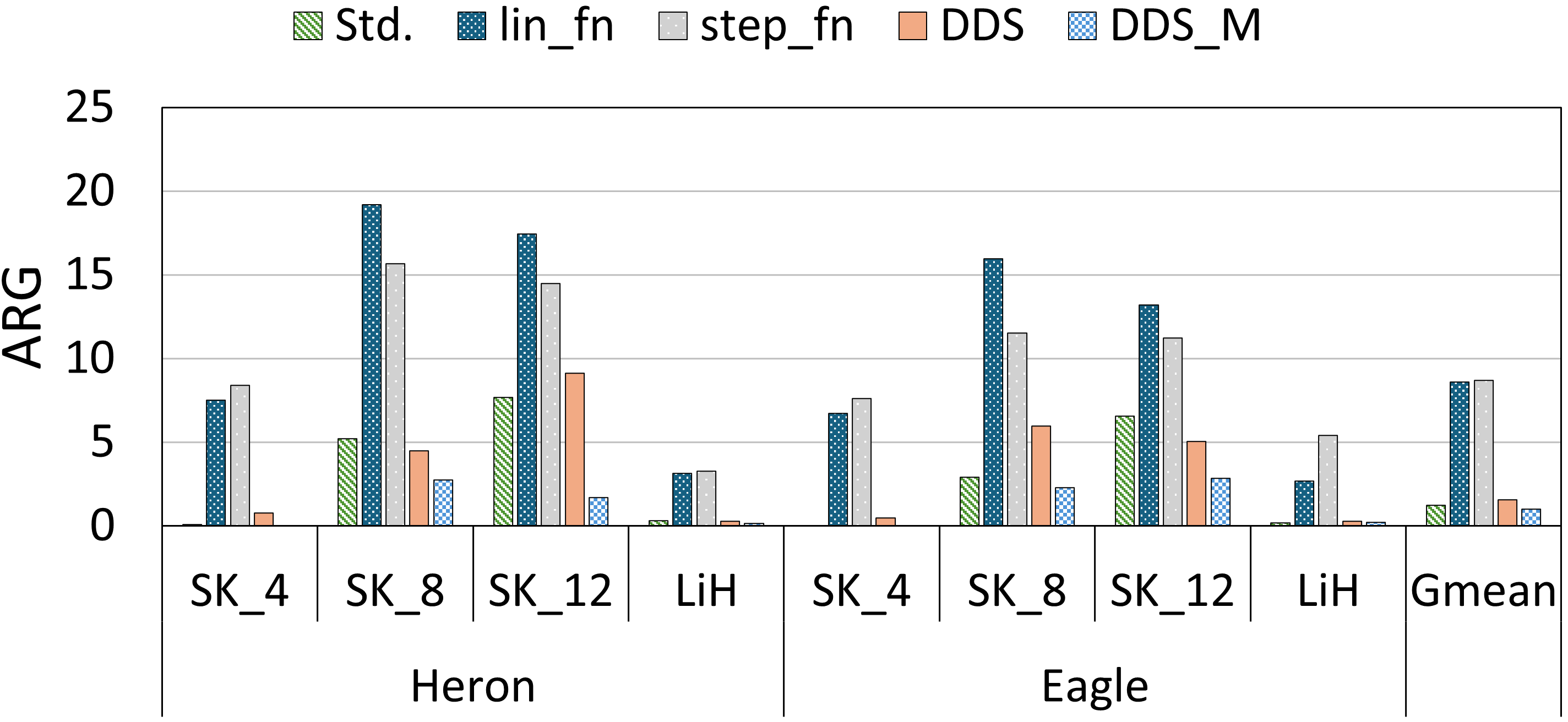}
  \caption{The ARG values for the training of Sherrington–Kirkpatrick (SK) QAOA model across different qubit counts and LiH VQE simulations under noisy condition. The qiskit-aer qasm simulator was used, and the native gate set along with error information from the $ibm\_marrakesh$ and $ibm\_yonsei$ processor, based on the Heron and Eagle architecture, were employed.}
  \label{fig:14}
\end{figure}

Figure \ref{fig:13} compares the total shot counts across different shot allocation methods, while Figure \ref{fig:14} illustrates the ARG performance under noisy conditions. The linear\_fn and step\_fn methods achieve significant reductions in total shot count, with average decreases of 87.13\% and 86.17\%, respectively, compared to the standard method. However, these methods exhibit substantial increases in ARG, by 599.19\% and 608.02\%, respectively. While linear\_fn and step\_fn effectively reduce shot counts even in noisy environments, they suffer from a more pronounced accuracy drop.

DDS achieves an average reduction in total shot count of 31.14\%, while increasing ARG by 26.92\% compared to the standard method. Although DDS requires more shots than linear\_fn and step\_fn, it still reduces the shot count compared to the standard approach, with ARG values 71.83\% and 65.98\% lower than those of linear\_fn and step\_fn, respectively. While DDS\_M uses more shots than DDS, it demonstrates better accuracy. Specifically, DDS\_M uses 4.30\% fewer shots than the standard method, while reducing ARG by 18.05\%. This indicates that DDS\_M ensures higher accuracy without increasing the overall shot count, even when more shots are required during specific iterations of the training process. This further highlights the effectiveness of determining shot counts based on the entropy of the quantum state's probability distribution. Under noisy conditions, details regarding the number of iterations, cost, and shot usage during the training process for QAOA simulations with the SK model and VQE simulations for LiH can be found in \ref{sec:Appendix-F}.

\subsubsection{\label{sec:Evaluation-B-3}Dynamic Shots on Entropy}

As discussed in Section-\ref{sec:Observation-C}, determining the number of shots based on entropy requires a constant parameter. As the number of qubits increases, entropy also increases, often requiring a smaller constant value to maintain the desired shot count. In the QAOA experiments presented above, we set this constant to 64 for 4-qubit circuits, 8 for 8-qubit circuits, and 2 for 12-qubit circuits. The constant value can be adjusted depending on both the number of qubits and the complexity of the quantum circuit. For each specific QAOA model and qubit count, we configured the constant $k$ to identify an optimal value.

\begin{figure}[b]
  \centering
  \includegraphics [width=\columnwidth] {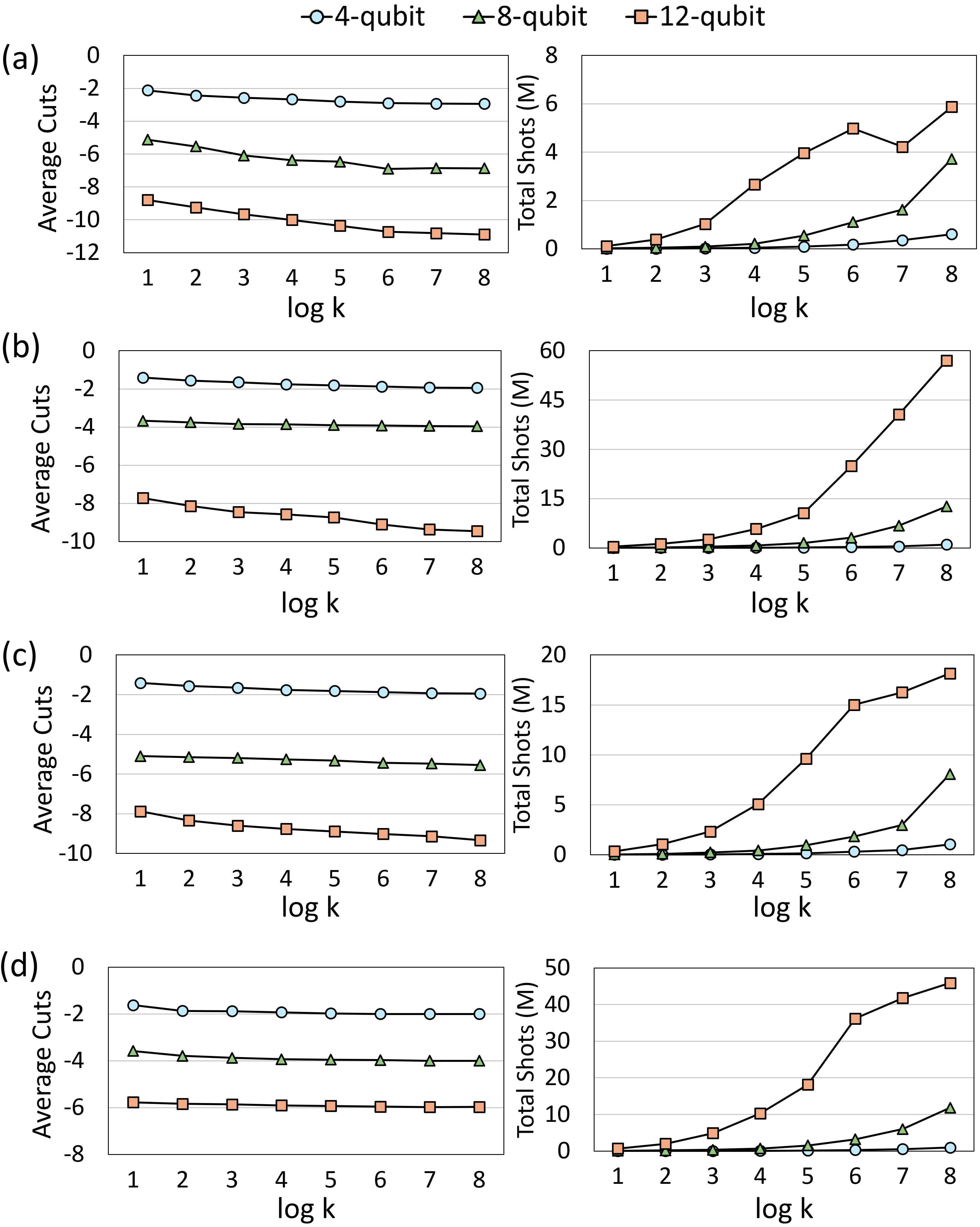}
  \caption{The average cuts and total shots for the QAOA training process of (a) PL, (b) BA, (c) WS, and (d) SK models.}
  \label{fig:15}
\end{figure}

Figure \ref{fig:15} shows the total number of shots and the average cuts for various QAOA models and qubit counts, with $k$ varied as powers of 2. In these experiments, the maximum shot count was increased to 100,000, compared to previous setups. For the Power-Law (PL) model, entropy is relatively low because the quantum state in the final iteration has a limited number of high-probability states. Therefore, the final cost is more sensitive to changes in $k$ compared to other models. In contrast, the Sherrington-Kirkpatrick (SK) model, which has the highest complexity, exhibits higher entropy. This results in a larger number of shots, as the higher entropy leads to a greater shot count, even for small values of $k$, ensuring sufficient sampling. The results show that the final cost for this model remains relatively stable across different values of $k$.

Examining the total shot count as a function of $k$ reveals that, as the number of qubits increases, the required shot count grows significantly. For circuits with a small qubit count, increasing $k$ to enhance accuracy has minimal impact on the total shot count. In contrast, for circuits with a large number of qubits, a high $k$ value can significantly increase the shot requirement. In such cases, adopting a smaller $k$ may provide a more efficient strategy, facilitating the effective management of shot resources while incurring only a slight trade-off in accuracy. This approach is particularly relevant for complex circuits, where the final cost remains largely unaffected by $k$, thereby reducing the risk of using a smaller constant.
\section{Conclusion}

Measurement-based quantum computers require repeated circuit executions to obtain the output distribution of the final quantum state. Dynamically adjusting the number of measurement shots per iteration can help reduce the overall execution time of VQA. Our observations show that, as the VQA training progresses, the information entropy of the probability distribution generally decreases. When entropy is low, fewer measurements are often sufficient to compute the cost function of VQA with adequate accuracy. To leverage this, we propose a Distribution-adaptive Dynamic Shot (DDS) method that adjusts the shot count dynamically based on the information entropy observed in the previous VQA iteration.

The DDS method reduces the required total shot count by 50.69\% compared to the conventional fixed-shot method \cite{phalak2023shot}, while maintaining similar accuracy and training epochs. Although our approach requires more shots than the tiered shot allocation method, it achieved a 63.34\% improvement in accuracy, demonstrating its capability to reduce the total number of shots while maintaining the accuracy of VQA. Under noisy conditions, DDS achieves an average 31.15\% reduction in the total shot count compared to the baseline, while maintaining a slight increase in accuracy. Moreover, compared to the tiered-shot approach, DDS provides a 68.91\% improvement in accuracy, even though it requires a higher shot count, demonstrating its robustness and effectiveness in noisy environments.

The DDS approach enables a significant reduction in the number of shots per iteration while maintaining the total number of VQA iterations. Moreover, DDS is applicable not only to current NISQ devices but also to future fault-tolerant quantum computers. As quantum hardware advances, achieving accurate probability distributions will entail substantial computational costs, particularly due to the large number of measurements required for tasks such as error correction and mitigation. DDS demonstrates its ability to enhance performance even under such computationally demanding conditions. Additionally, the integration of DDS with current efficient VQA algorithms \cite{zhao2024enhancing, eddins2022doubling, huggins2022unbiasing, gujarati2023quantum, wu2024variational, cimini2024variational, rahman2024fine} can create a synergistic effect, further improving and accelerating the practical applicability of VQA algorithms.

In cloud-based quantum computing systems, multiple tenants frequently submit jobs, resulting in high job volumes that lead to extended wait times for receiving results. For variational quantum algorithms, each iteration must be queued sequentially, as classically updated parameters are required for subsequent quantum circuits. Consequently, iterations are often delayed, as they cannot commence immediately after the previous one. Reducing the number of measurement shots can alleviate queuing delays for all tenants and decrease classical post-processing overhead. Furthermore, the DDS approach presents a viable method for enhancing cost efficiency in cloud-based quantum computing \cite{wang2024qoncord}. By minimizing the total number of shots required, DDS significantly lowers cloud usage costs, particularly in scenarios where access to high-precision quantum devices is scarce.

\begin{acknowledgments}
This work was supported by Hyundai Motor Company (HMC). The authors thank Dr. Seunghyo Noh at HMC for enlightening discussions.
\end{acknowledgments}

\nocite{*}

\bibliography{mainbib}

\appendix
\section{\label{sec:Appendix-A}Measurement in Quantum Computing}
In quantum computing, circuit execution concludes by measuring the final state of the qubits. A quantum state can be represented as a superposition of basis states, and each measurement yields a single basis state. To construct the final probability distribution over these basis states, numerous measurements (referred to as "shots") are required. The number of shots is determined by the programmer, and the maximum number of shots, along with default settings, is specified in the quantum system's documentation. For each measurement, all qubits collapse to one of the basis states. After this collapse (typically to the computational basis), the circuit must be reconstructed to perform subsequent measurements.

The number of shots required can vary depending on the desired accuracy and the quantum system's noise characteristics. Reducing the number of shots while maintaining accuracy is a key challenge, particularly in noisy intermediate-scale quantum devices. Techniques like dynamic shot allocation can optimize shot counts while preserving computational fidelity. By carefully balancing shot numbers with measurement strategies, the efficiency of quantum computations can be improved, maximizing the effective use of available quantum resources.

\section{\label{sec:Appendix-B}Variational Quantum Algorithms}
VQA \cite{cerezo2021variational, biamonte2017quantum, rebentrost2014quantum, mcclean2016theory, havlivcek2019supervised} is a quantum algorithm that minimizes the value of a given function, such as finding the ground state energy of a molecule \cite{peruzzo2014variational, kandala2017hardware} or solving the maximum cut problem in QAOA \cite{blekos2024review, zhou2020quantum}. The two main components of VQAs are the parameterized quantum part, which executes the quantum circuit that reflects the loss function, and the classical optimization part, which calculates the cost value from the results and optimizes the parameters of the quantum circuit. The quantum part includes preparing the quantum circuit and executing it to generate measurement outcomes. An ansatz is used to define the VQA circuit, representing the function to be solved.

An ansatz is a parameterized circuit composed of gates that can span the entire Bloch sphere for a single qubit or cover all norm vectors for multiple qubits. This ansatz defines the quantum circuit that represents the problem to be solved. In the circuit preparation process, the programmer initially sets the ansatz and its parameters, creating the quantum circuit. The circuit is then decomposed into native gates and executed on the quantum processor. After measuring the quantum state, the classical computer processes the measurement results to calculate the cost value and updates the quantum circuit's parameters to minimize this cost. With the updated parameters, the ansatz is re-prepared, and the next iteration is executed. This process is repeated until the classical optimizer concludes that the cost value is minimized.
\section{\label{sec:Appendix-C} Entropy of Random Number Generation Circuit}

In the context of quantum computing, the entropy of a system represents the uncertainty or randomness associated with its quantum state distribution of possible outcomes.  
The random number generation circuit is particularly relevant in this regard, as it produces quantum states with varying entropy levels based on the number of qubits and the nature of the transformations applied.

Entropy, specifically \emph{Shannon entropy}, quantifies the uncertainty of a probability distribution \( P = \{p_1, p_2, \dots, p_n\} \), where \( p_i \) represents the probability of observing the \( i \)-th outcome among all possible outcomes.
The formula for Shannon entropy is expressed as follows:
\begin{equation}
    H(P) = - \sum_{i=1}^n p_i \log_2 p_i,
\end{equation}
where \( H(P) \) is the entropy, \( n \) is the total number of possible outcomes (determined by the number of qubits), and \( p_i \) is the probability of the \( i \)-th basis state or outcome.

For a random number generation circuit, the probability distribution \( P \) is derived from the circuit's measurements.
A uniform distribution of \( p_i = \frac{1}{n} \) for all \( i \) results in the maximum entropy for n-qubit quantum circuits:
\begin{equation}
    H_{\text{rand}} = - \sum_{i=1}^n \frac{1}{n} \log_2 \frac{1}{n} =  \log_2 n.
\end{equation}

Conversely, non-uniform distributions result in lower entropy values, reflecting reduced randomness.
In these cases, certain outcomes become more probable, which increases the system’s overall unpredictability and indicates a distinct bias in the distribution of quantum states.
\section{\label{sec:Appendix-D} Performance Comparison of Shot Approaches}

\begin{table*}[t]
\centering
\renewcommand{\arraystretch}{1}
\renewcommand{\tabcolsep}{2.2mm}
\caption{Performance comparison of various shot allocation methods, including Power-Law (PL), Barabási–Albert (BA), Watts–Strogatz (WS), and Sherrington–Kirkpatrick (SK), is presented across a range of qubit sizes to evaluate the effectiveness of these approaches under varying quantum circuit complexities. The table provides a comprehensive overview, detailing the average cuts after all iterations, the total number of iterations (epochs) performed, and the average shots per iteration ($S_{avg}$), which is calculated by dividing the total number of shots by the total number of iterations. The shot allocation methods considered include Std., a conventional fixed-shot approach utilizing 1,024 shots per iteration; lin\_fn, a tiered shot allocation method based on a linear function that adjusts shot counts progressively; step\_fn, a tiered shot allocation method based on a step function that reduces shots in discrete intervals; and DDS, the proposed entropy-based dynamic shot allocation method, which adapts the number of shots dynamically based on the entropy of the quantum state throughout the optimization process.}
\begin{tabular}{|c|c|c|c|c|c|c|c|c|c|c|c|c|}
\hline
\multirow{2}{*}{Application} & \multicolumn{4}{c|}{4-qubit PL QAOA} & \multicolumn{4}{c|}{8-qubit PL QAOA} & \multicolumn{4}{c|}{12-qubit PL QAOA} \\ \cline{2-13} 
                             & Std. & lin\_fn & step\_fn & DDS & Std. & lin\_fn & step\_fn & DDS & Std. & lin\_fn & step\_fn & DDS \\ \hline
$S_{avg}$                    & 1,024 & 159.05 & 164.08 & 370.3 & 1,024 & 159.45 & 157.36 & 431.4 & 1,024 & 160.26 & 157.36 & 565 \\ \hline
Iterations                   & 185 & 350 & 369 & 193 & 442 & 349 & 387 & 385 & 450 & 347 & 387 & 395 \\ \hline
Average Cuts                   & -2.68 & -2.49 & -2.42 & -2.64 & -6.77 & -5.73 & -6.10 & -6.80 & -9.98 & -8.98 & -8.77 & -10.24 \\ \hline

\multirow{2}{*}{Application} & \multicolumn{4}{c|}{4-qubit BA QAOA} & \multicolumn{4}{c|}{8-qubit BA QAOA} & \multicolumn{4}{c|}{12-qubit BA QAOA} \\ \cline{2-13} 
                             & Std. & lin\_fn & step\_fn & DDS & Std. & lin\_fn & step\_fn & DDS & Std. & lin\_fn & step\_fn & DDS \\ \hline
$S_{avg}$                    & 1,024 & 163.16 & 164.47 & 577.6 & 1,024 & 152.60 & 174.58 & 745.2 & 1,024 & 153.32 & 165.66 & 431.13 \\ \hline
Iterations                   & 392 & 340 & 368 & 420 & 461 & 367 & 344 & 390 & 455 & 365 & 365 & 406 \\ \hline
Average Cuts                   & -1.89 & -1.65 & -1.67 & -1.85 & -3.86 & -3.13 & -3.23 & -3.89 & -8.03 & -6.04 & -6.24 & -7.92 \\ \hline

\multirow{2}{*}{Application} & \multicolumn{4}{c|}{4-qubit WS QAOA} & \multicolumn{4}{c|}{8-qubit WS QAOA} & \multicolumn{4}{c|}{12-qubit WS QAOA} \\ \cline{2-13} 
                             & Std. & lin\_fn & step\_fn & DDS & Std. & lin\_fn & step\_fn & DDS & Std. & lin\_fn & step\_fn & DDS \\ \hline
$S_{avg}$                    & 1,024 & 163.16 & 164.47 & 577.6 & 1,024 & 162.32 & 149.63 & 87.1 & 1,024 & 145.08 & 158.80 & 391.3 \\ \hline
Iterations                   & 392 & 340 & 368 & 420 & 374 & 342 & 410 & 362 & 432 & 389 & 383 & 435 \\ \hline
Average Cuts                   & -1.89 & -1.65 & -1.67 & -1.85 & -4.94 & -4.05 & -3.90 & -4.88 & -7.76 & -6.57 & -6.84 & -7.65 \\ \hline

\multirow{2}{*}{Application} & \multicolumn{4}{c|}{4-qubit SK QAOA} & \multicolumn{4}{c|}{8-qubit SK QAOA} & \multicolumn{4}{c|}{12-qubit SK QAOA} \\ \cline{2-13} 
                             & Std. & lin\_fn & step\_fn & DDS & Std. & lin\_fn & step\_fn & DDS & Std. & lin\_fn & step\_fn & DDS \\ \hline
$S_{avg}$                    & 1,024 & 168.41 & 171.92 & 534.5 & 1,024 & 141.33 & 169.77 & 688.4 & 1,024 & 151.16 & 161.01 & 887.9 \\ \hline
Iterations                   & 360 & 328 & 350 & 371 & 409 & 401 & 355 & 377 & 422 & 371 & 377 & 420 \\ \hline
Average Cuts                   & -2.00 & -1.84 & -1.88 & -2.00 & -3.86 & -3.63 & -3.36 & -3.82 & -5.72 & -4.94 & -4.94 & -5.79 \\ \hline

\end{tabular}
\label{Table:t1}
\end{table*}

\begin{table*}[t]
\centering
\renewcommand{\arraystretch}{1}
\renewcommand{\tabcolsep}{2.3mm}
\caption{Performance comparison of shot allocation methods applied to VQE simulations for H$_{2}$, LiH, and BeH$_{2}$ molecules. In the VQE simulations, the energy calculated as the expectation value corresponds to the electronic ground-state energy. To obtain the total ground-state energy, the nuclear repulsion energy was added. All energy values are reported in Hartree units.}
\begin{tabular}{|c|c|c|c|c|c|c|c|c|c|c|c|c|}
\hline
\multirow{2}{*}{Application} & \multicolumn{4}{c|}{H$_{2}$ VQE} & \multicolumn{4}{c|}{LiH VQE} & \multicolumn{4}{c|}{BeH$_{2}$ VQE} \\ \cline{2-13} 
                             & Std. & lin\_fn & step\_fn & DDS & Std. & lin\_fn & step\_fn & DDS & Std. & lin\_fn & step\_fn & DDS \\ \hline
$S_{avg}$                    & 1,024 & 848.57 & 897.06 & 137.7 & 1,024 & 89.63 & 85.30 & 570.3 & 1,024 & 68.51 & 73.00 & 928.32 \\ \hline
Iterations                   & 40 & 36 & 35 & 37 & 659 & 698 & 813 & 658 & 1,001 & 1,001 & 1,001 & 1,001 \\ \hline
Energy (E$_h$)                  & -1.12 & -1.12 & -1.13 & -1.12 & -7.83 & -7.58 & -7.76 & -7.86 & -15.02 & -14.96 & -14.69 & -15.02 \\ \hline
\end{tabular}
\label{Table:t2}
\end{table*}

Table \ref{Table:t1} and Table \ref{Table:t2} present the final cost, total iterations (i.e., epochs), and the average number of shots per iteration for different shot allocation methods used in the training of each VQA model. The total number of iterations (denoted as $Iterations$) excludes the final common 1,024-shot evaluation described in Section \ref{sec:Evaluation-B}. Additionally, $S_{avg}$ represents the total number of shots used throughout the training process, divided by the total number of iterations, providing an estimate of shot efficiency for each method. The shot allocation strategies vary in how they distribute shots across training iterations, affecting both training efficiency and accuracy.

For each shot allocation approach, there are the fixed shot method, tiered shot allocation method from previous work, and DDS. The ``Std." method represents a standard shot approach, maintaining a constant 1,024 shots per iteration without adjustments. The ``lin\_fn" and ``step\_fn" methods are tiered shot allocation strategies. The lin\_fn starts with 1,000 shots and reduces the number of shots linearly over iterations, while the step\_fn decreases the shot count by 10 every 10 iterations. Lastly, the ``DDS" method dynamically determines the number of shots based on the entropy of the probability distribution from the previous iteration, as detailed in the paper.

For the QAOA models (PL, BA, WS, and SK), simulations were conducted with qubit counts of 4, 8, and 12. For the VQE models, molecular systems including H$_{2}$, LiH, and BeH$_{2}$ were simulated. Across all VQA applications and varying qubit numbers, DDS demonstrated a significant reduction in the required number of shots compared to the standard method, while maintaining a similar total number of iterations. Specifically, for QAOA, the average reduction in shot count was 50.69\%, and for VQE, it was 47.09\%. Notably, the final cost values showed negligible differences between DDS and the standard method, indicating that DDS achieves comparable accuracy with fewer shots. When comparing the final cost to the ideal value using the ARG metric, an average increase of only 6.03\% was observed for QAOA and VQE. Interestingly, for VQE, the ARG value actually decreased by 24.39\%, indicating that DDS may offer an efficient approach to cost estimation with fewer shots.

The lin\_fn and step\_fn methods reduce the shot count by 85.49\% and 84.24\%, respectively, compared to the standard method. However, these methods incur significant accuracy losses, as indicated by ARG increases of 241.18\% and 230.73\%, respectively. In contrast, DDS dynamically adjusts the number of shots, minimizing accuracy degradation while still requiring fewer shots than the conventional fixed-shot approach. Although DDS requires more shots than the tiered shot allocation methods, it results in an overall ARG reduction of 63.34\%, demonstrating a substantial improvement in accuracy.
\section{\label{sec:Appendix-E} Evaluation of Shot Allocation Methods Against the Standard}

\begin{figure}[!htbp]
  \centering 
  \includegraphics [width=0.75\columnwidth] {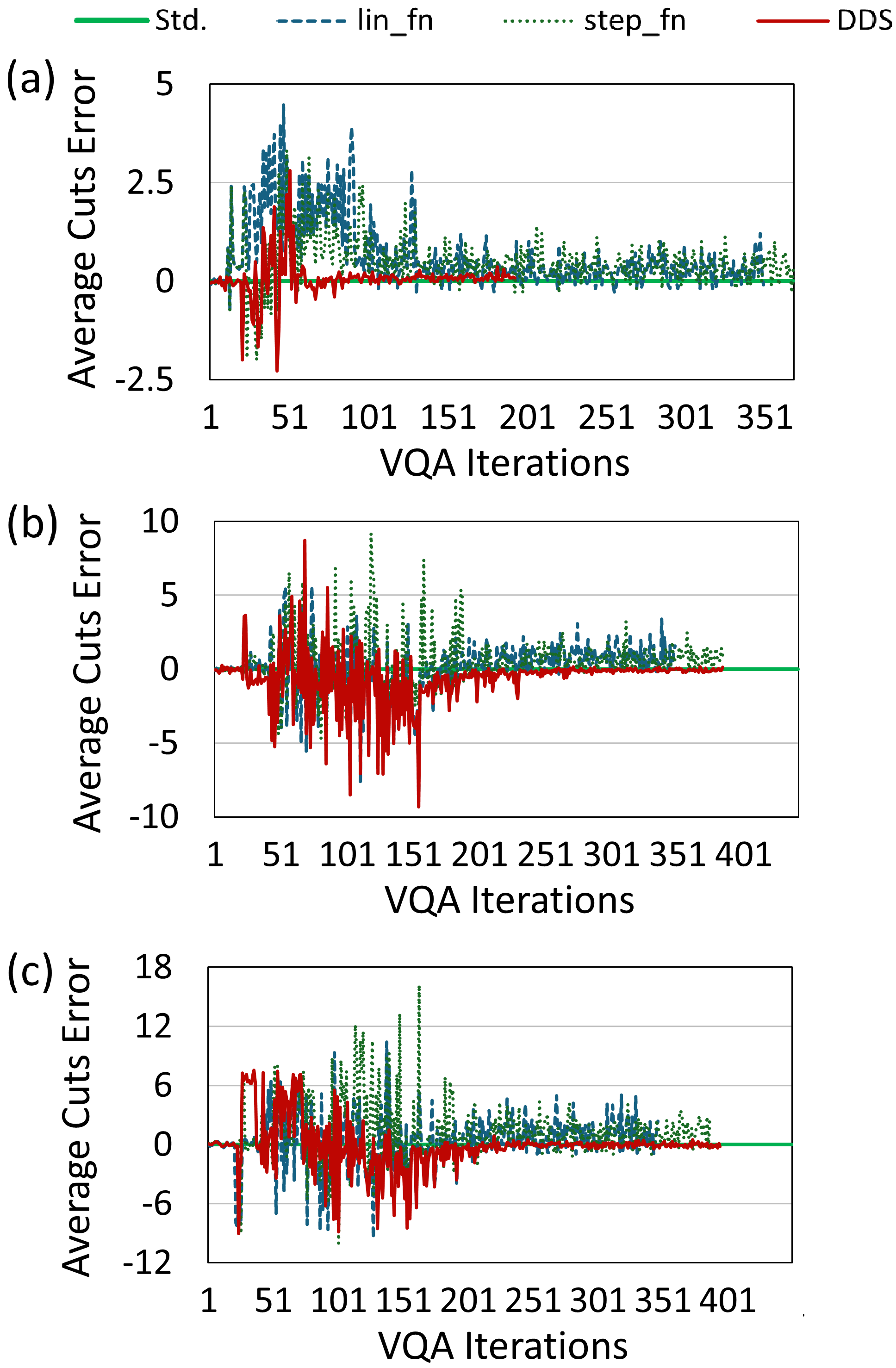}
  \caption{The average cuts error per iteration against standard shot allocation method during the PL model QAOA simulation for (a) 4 qubits, (b) 8 qubits, and (c) 12 qubits.}
  \label{fig:16}
\end{figure}

\begin{figure}[!htbp]
  \centering 
  \includegraphics [width=0.75\columnwidth] {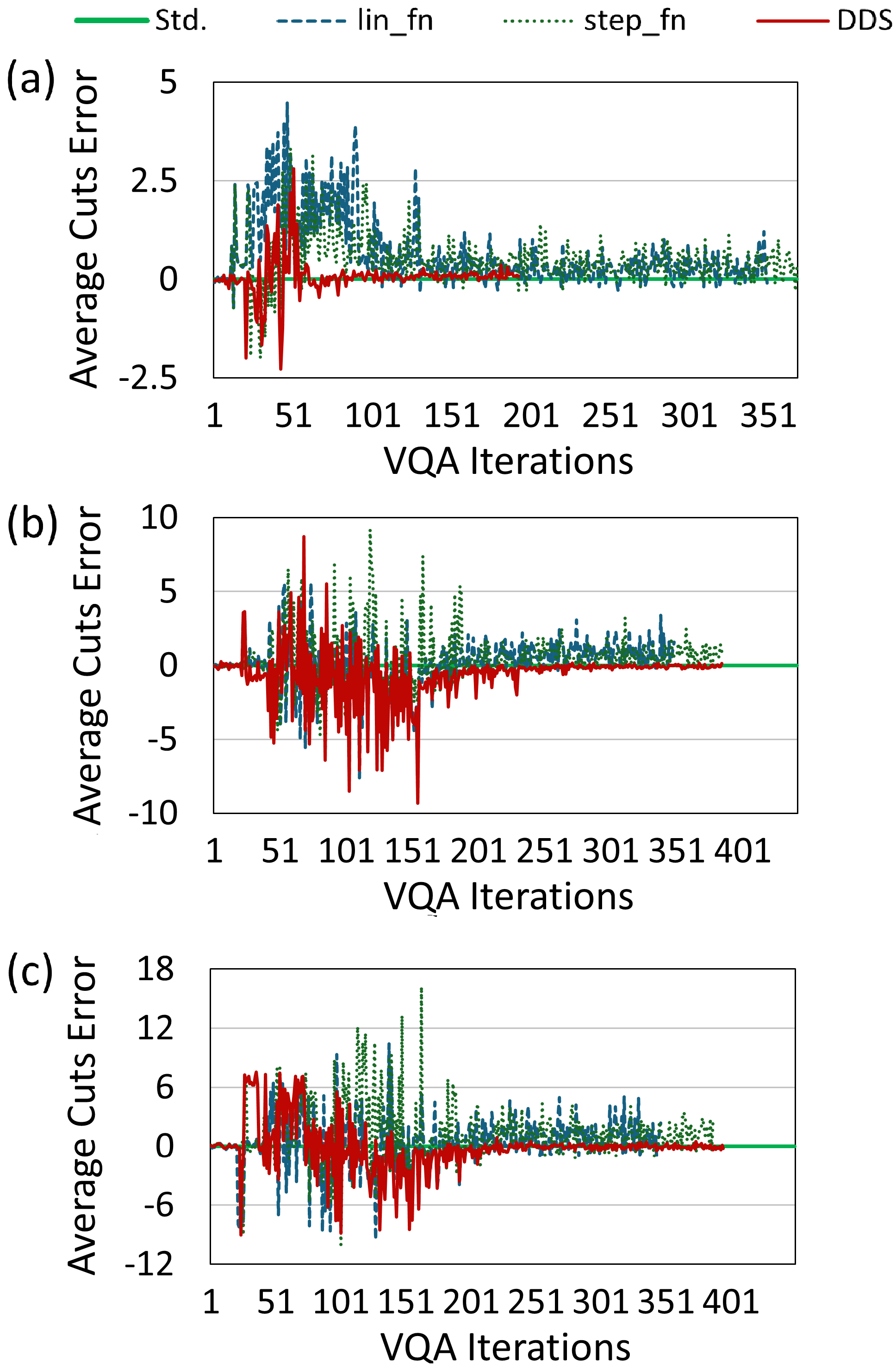}
  \caption{The average cuts error per iteration against standard shot allocation method during the BA model QAOA simulation for (a) 4 qubits, (b) 8 qubits, and (c) 12 qubits.}
  \label{fig:17}
\end{figure}

\begin{figure}[!htbp]
  \centering 
  \includegraphics [width=0.75\columnwidth] {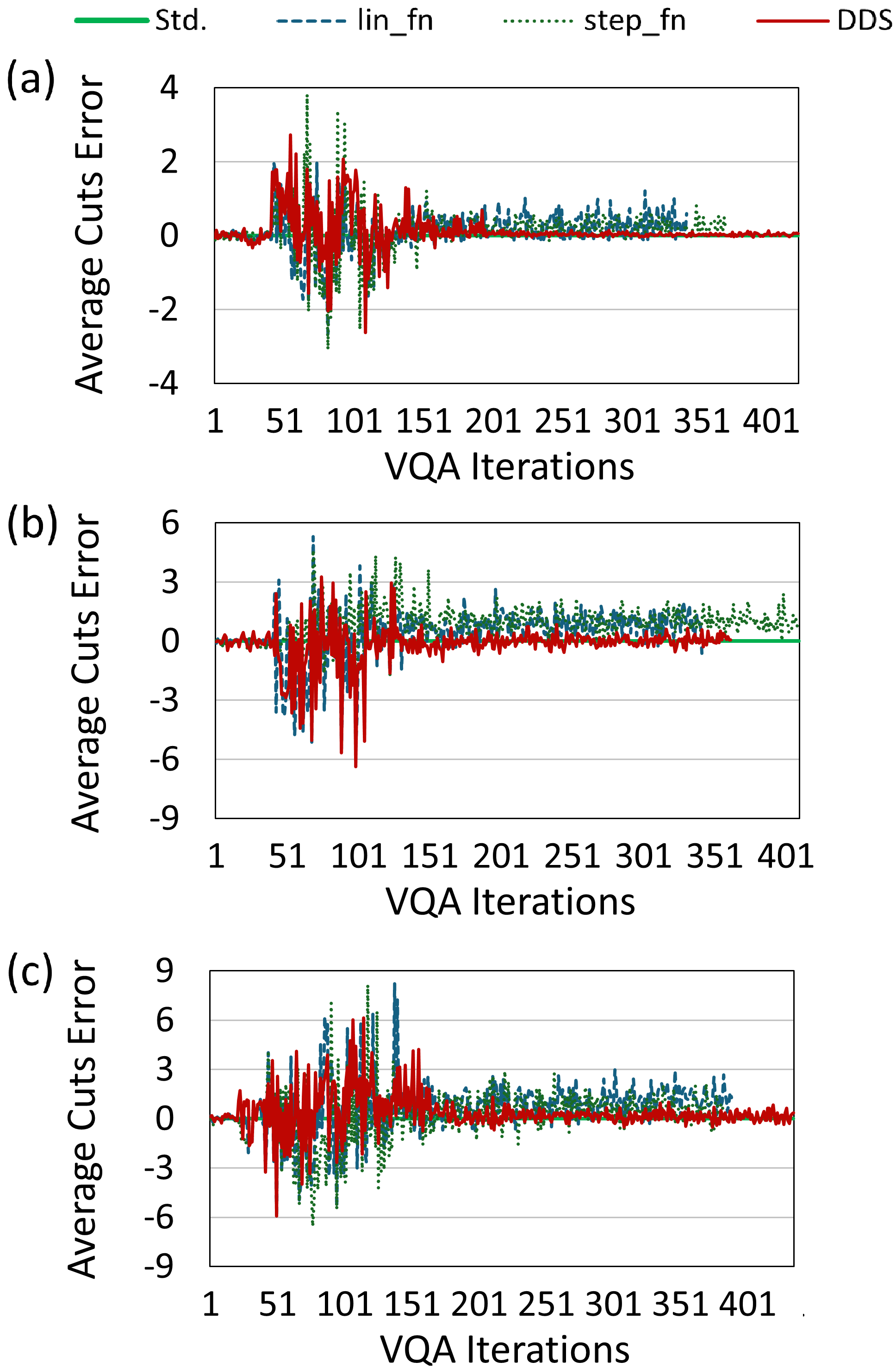}
  \caption{The average cuts error per iteration against standard shot allocation method during the WS model QAOA simulation for (a) 4 qubits, (b) 8 qubits, and (c) 12 qubits.}
  \label{fig:18}
\end{figure}

\begin{figure}[!htbp]
  \centering 
  \includegraphics [width=0.75\columnwidth] {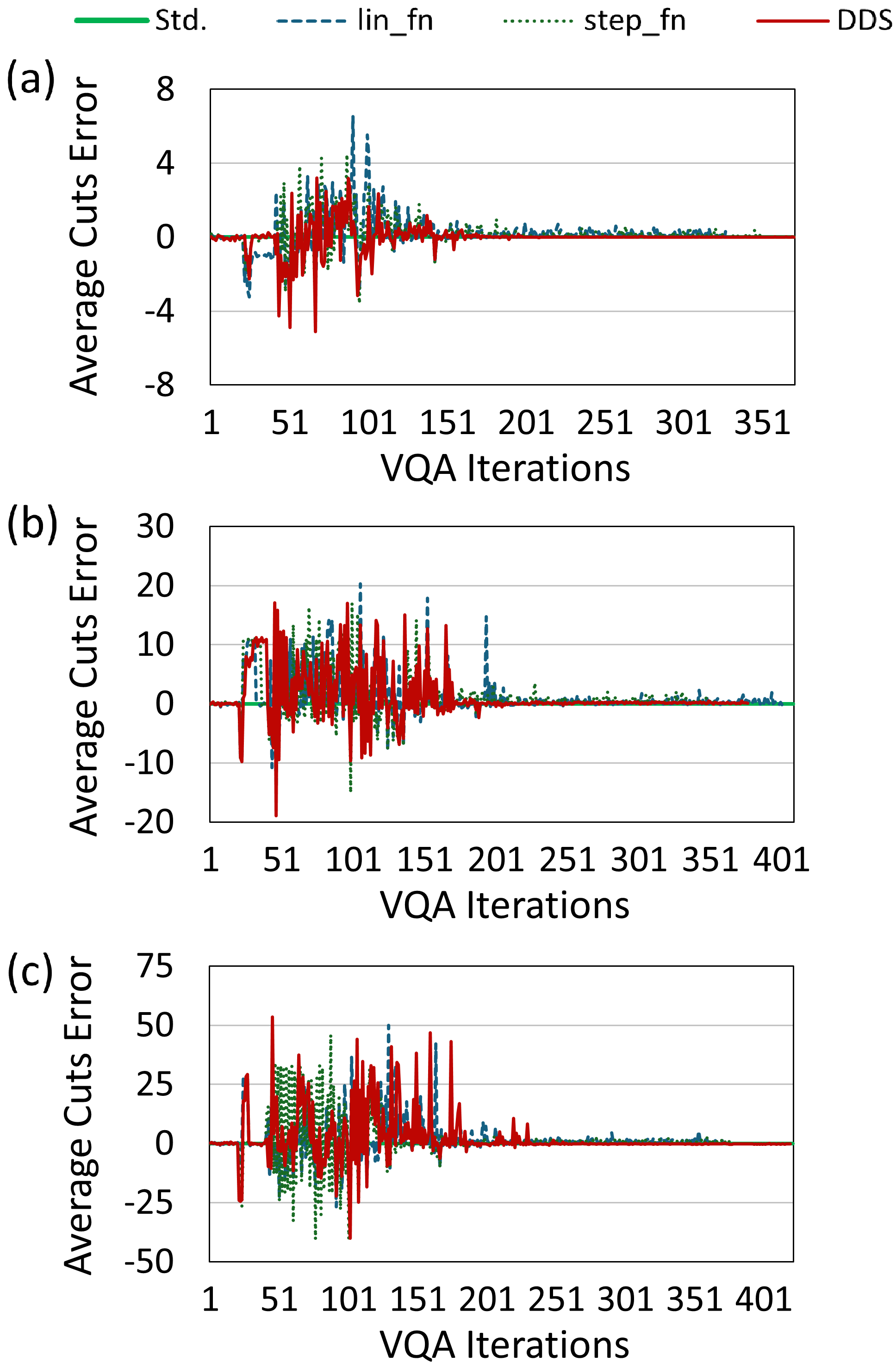}
  \caption{The average cuts error per iteration against standard shot allocation method during the SK model QAOA simulation for (a) 4 qubits, (b) 8 qubits, and (c) 12 qubits.}
  \label{fig:19}
\end{figure}

\begin{figure}[!htbp]
  \centering
  \includegraphics [width=0.75\columnwidth] {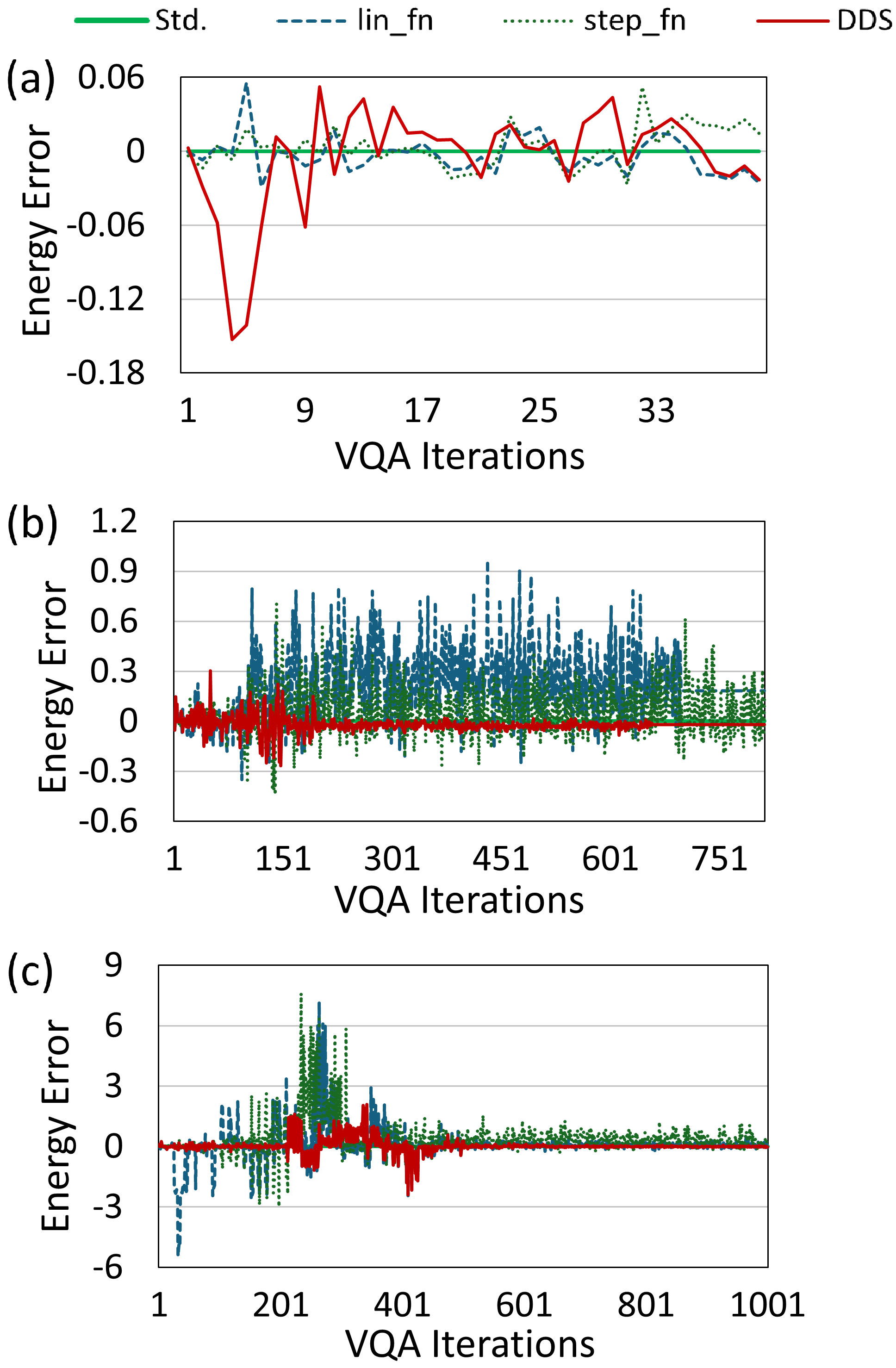}
  \caption{The energy error per iteration against standard shot allocation method for the VQE simulation of (a) H$_{2}$, (b) LiH, and (c) BeH$_{2}$.}
  \label{fig:20}
\end{figure}

When employing different shot allocation methods in VQA simulations, the final cost typically converges to a specific value as the iterations progress. This value represents the average cuts in the case of QAOA and the energy in the case of VQE. It is considered more effective if the cost converges faster to a lower value and exhibits minimal fluctuations during the convergence process. Conversely, if the cost fails to converge accurately in later iterations and exhibits significant fluctuations, it may hinder the accurate measurement of the final value.

Figures \ref{fig:16}, \ref{fig:17}, \ref{fig:18}, and \ref{fig:19} present the average cuts calculated at each iteration for QAOA simulations applied to the PL, BA, WS, and SK models, using various shot allocation methods compared to the standard shot allocation approach. The results indicate that both linear\_fn and step\_fn exhibit slower convergence compared to DDS across all QAOA models. Furthermore, these methods do not converge to a steady value but instead display oscillatory behavior throughout the iterations, persisting until the final iteration.

Figure \ref{fig:20} presents the energy variations at each iteration during VQE simulations of H$_2$, LiH, and BeH$_2$ using different shot allocation methods compared to the standard fixed shot allocation approach. In the case of H$_2$, due to its small number of qubits, the total number of iterations is relatively low, and the differences between shot allocation methods are not particularly pronounced. However, for LiH and BeH$_2$, it is evident that the linear\_fn and step\_fn methods exhibit slower convergence and display oscillatory behavior until the final iteration.

Both QAOA and VQE utilize fewer shots compared to the standard fixed shot method when applying the shot allocation approach. Notably, in the case of DDS, there are instances where the computed cost is lower than that of the standard method. A lower calculated cost during iterations indicates the possibility for the optimization process to converge more quickly to the lowest cost. As observed in the figures, better training performance is indicated when the error approaches a negative value. DDS demonstrates the potential for improved convergence while requiring fewer shots than the fixed shot allocation method.
\section{\label{sec:Appendix-F} Dynamic Shot Allocation in Noisy Simulator}

\begin{figure}[!htbp]
  \centering
  \includegraphics [width=\columnwidth] {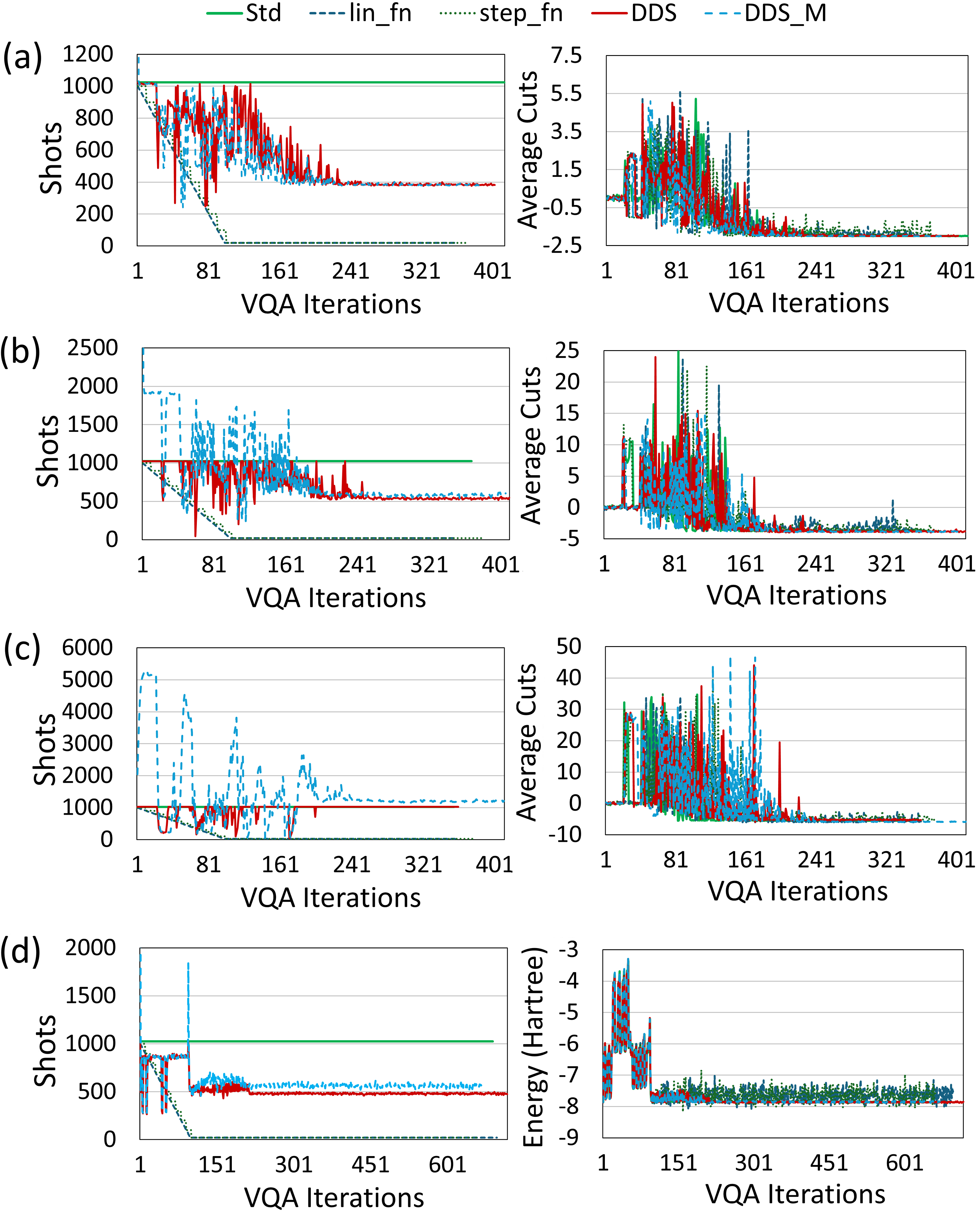}
  \caption{The number of shots and average cuts per iteration during the SK model QAOA training for (a) 4 qubits, (b) 8 qubits, and (c) 12 qubits, together with the number of shots and the energy of each iteration for the VQE training process of (d) LiH, using different shot allocation methods under noisy condition. The qiskit-aer qasm simulator was used, and the native gate set along with error information from the $ibm\_marrakesh$ processor, based on the Heron architecture, was employed.}
  \label{fig:21}
\end{figure}

\begin{figure}[!htbp]
  \centering
  \includegraphics [width=\columnwidth] {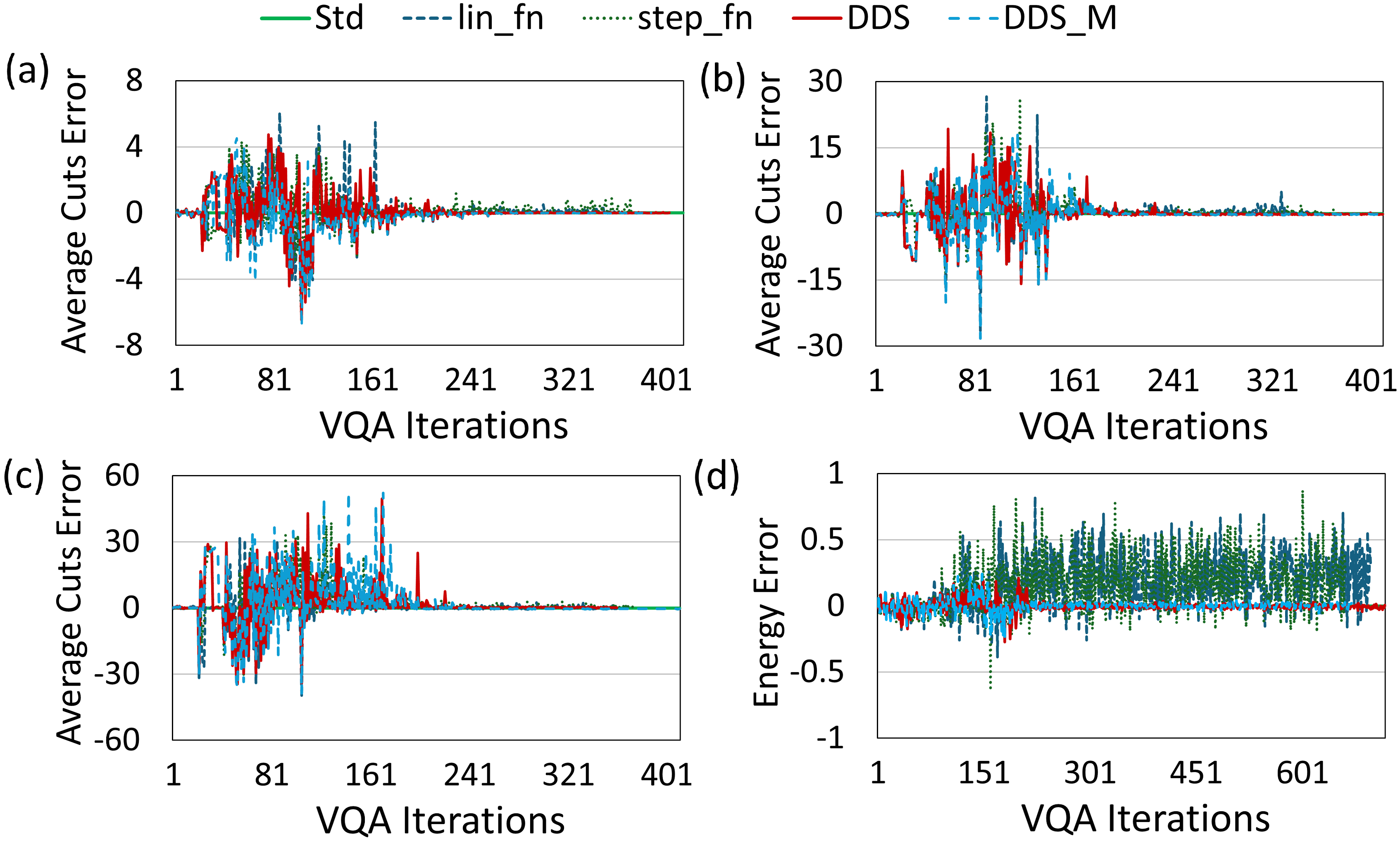}
  \caption{The average cuts error per iteration against the standard shot allocation method during the SK model QAOA simulation for (a) 4 qubits, (b) 8 qubits, and (c) 12 qubits, together with the energy error per iteration against the standard shot allocation method for the VQE simulation of (d) LiH. The qiskit-aer qasm simulator was used, and the native gate set along with error information from the $ibm\_marrakesh$ processor, based on the Heron architecture, was employed.}
  \label{fig:22}
\end{figure}

\begin{figure}[!htbp]
  \centering
  \includegraphics [width=\columnwidth] {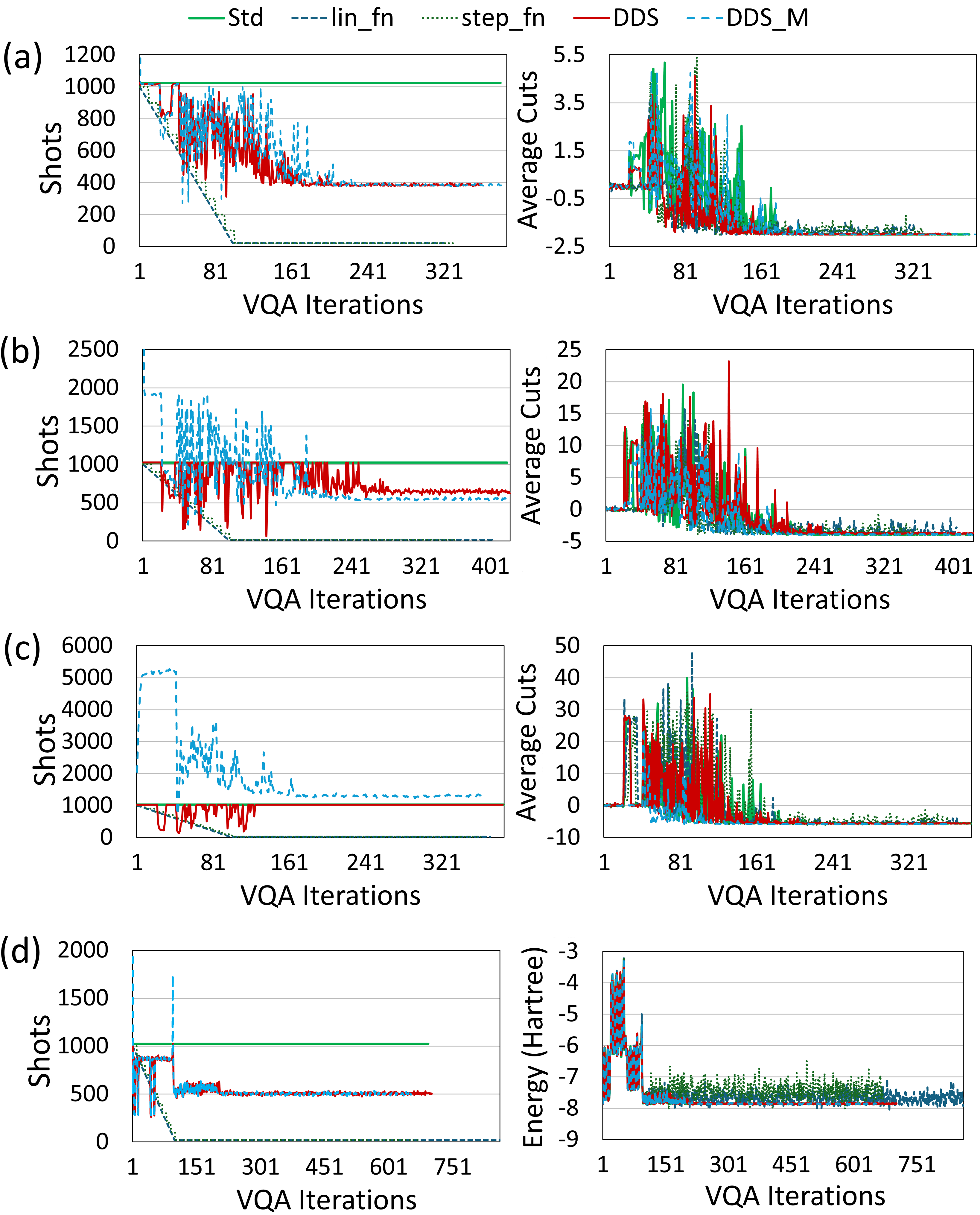}
  \caption{The number of shots and average cuts per iteration during the SK model QAOA training for (a) 4 qubits, (b) 8 qubits, and (c) 12 qubits, together with the number of shots and the energy of each iteration for the VQE training process of (d) LiH, using different shot allocation methods under noisy condition. The qiskit-aer qasm simulator was used, and the native gate set along with error information from the $ibm\_yonsei$ processor, based on the Eagle architecture, was employed.}
  \label{fig:23}
\end{figure}

\begin{figure}[!htbp]
  \centering
  \includegraphics [width=\columnwidth] {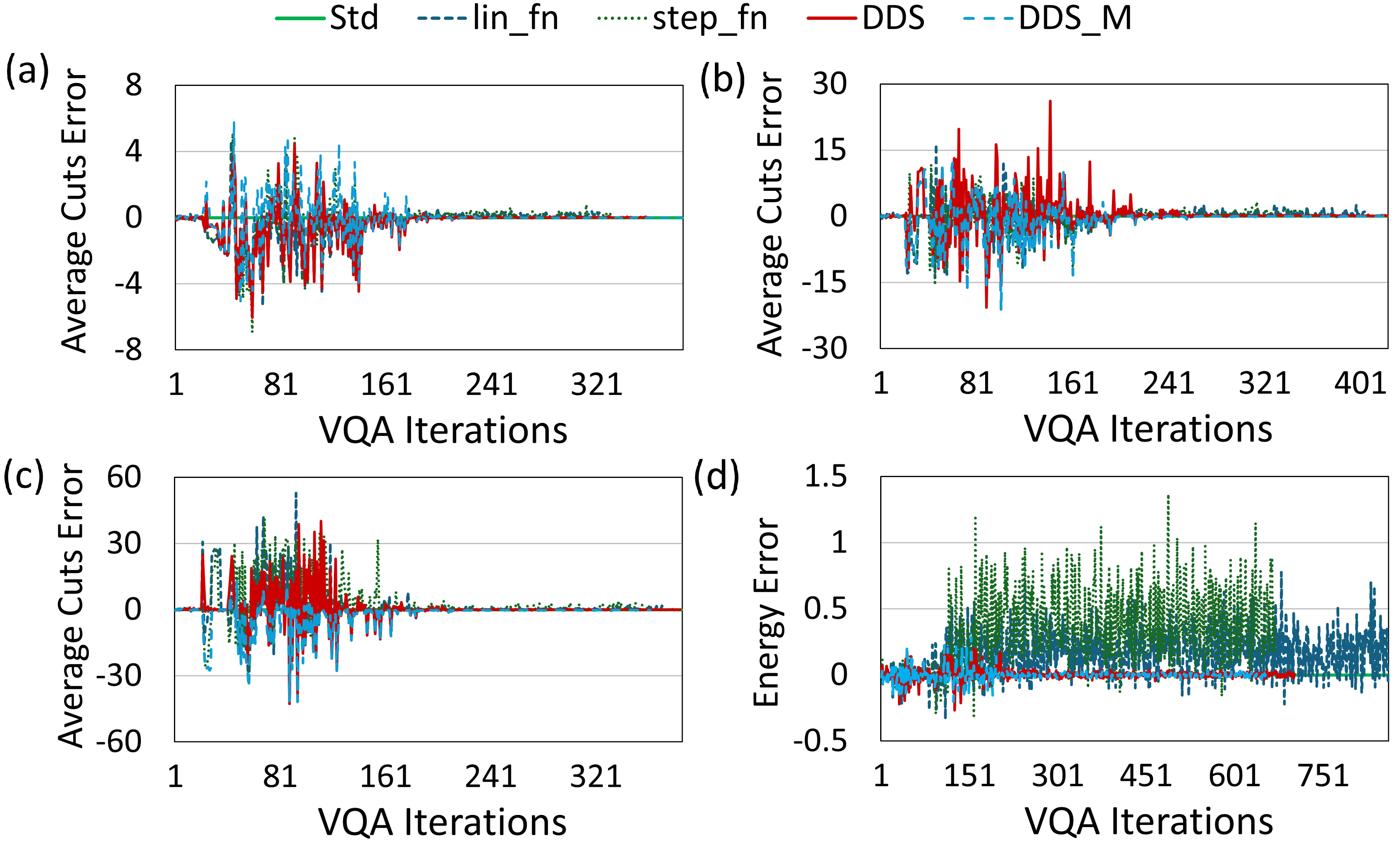}
  \caption{The average cuts error per iteration against the standard shot allocation method during the SK model QAOA simulation for (a) 4 qubits, (b) 8 qubits, and (c) 12 qubits, together with the energy error per iteration against the standard shot allocation method for the VQE simulation of (d) LiH. The qiskit-aer qasm simulator was used, and the native gate set along with error information from the $ibm\_yonsei$ processor, based on the Eagle architecture, was employed.}
  \label{fig:24}
\end{figure}

\begin{table*}[t]
\centering
\renewcommand{\arraystretch}{1}
\renewcommand{\tabcolsep}{0.5mm}
\caption{Performance comparison of various shot allocation methods under noisy condition, Sherrington–Kirkpatrick (SK) is presented across a range of qubit sizes to evaluate the effectiveness of these approaches under varying quantum circuit complexities. For each quantum processor architecture (Heron and Eagle), the ibm\_marrakesh and ibm\_yonsei processors are used to provide error information for the noise simulator. The table details the average cuts after all iterations, the total number of iterations (epochs) performed, and the average shots per iteration ($S_{avg}$). The shot allocation methods considered include Std., lin\_fn, step\_fn, DDS, and DDS\_M. DDS\_M is the DDS method that does not have the 1,024-shot limit. }
\begin{tabular}{|c|c|c|c|c|c|c|c|c|c|c|c|c|c|c|c|}
\hline

QPU Arch. & \multicolumn{15}{c|}{Heron} \\ \hline
\multirow{2}{*}{Application} & \multicolumn{5}{c|}{4-qubit SK QAOA} & \multicolumn{5}{c|}{8-qubit SK QAOA} & \multicolumn{5}{c|}{12-qubit SK QAOA} \\ \cline{2-16} 
                             & Std. & lin\_fn & step\_fn & DDS & DDS\_M & Std. & lin\_fn & step\_fn & DDS & DDS\_M & Std. & lin\_fn & step\_fn & DDS & DDS\_M \\ \hline
$S_{avg}$                    & 1,024 & 155.56 & 162.14 & 551.34 & 687.35 & 1,024 & 159.05 & 159.16 & 703.74 & 861.09 & 1,024 & 155.18 & 161.39 & 945.89 & 1,471.97 \\ \hline
Iterations                   & 414 & 359 & 374 & 403 & 388 & 368 & 350 & 382 & 410 & 406 & 355 & 360 & 376 & 360 & 411 \\ \hline
Average Cuts                   & -1.998 & -1.85 & -1.83 & -1.98 & -2.00 & -3.79 & -3.23 & -3.37 & -3.82 & -3.89 & -5.52 & -4.93 & -5.11 & -5.43 & -5.88 \\ \hline

QPU Arch. & \multicolumn{15}{c|}{Eagle} \\ \hline
\multirow{2}{*}{Application} & \multicolumn{5}{c|}{4-qubit SK QAOA} & \multicolumn{5}{c|}{8-qubit SK QAOA} & \multicolumn{5}{c|}{12-qubit SK QAOA} \\ \cline{2-16} 
                             & Std. & lin\_fn & step\_fn & DDS & DDS\_M & Std. & lin\_fn & step\_fn & DDS & DDS\_M & Std. & lin\_fn & step\_fn & DDS & DDS\_M \\ \hline
$S_{avg}$                    & 1,024 & 170.71 & 179.70 & 532.92 & 720.47 & 1,024 & 140.42 & 168.10 & 775.18 & 832.68 & 1,024 & 150.81 & 166.87 & 961.62 & 1,959.25 \\ \hline
Iterations                   & 380 & 323 & 333 & 360 & 386 & 420 & 404 & 359 & 423 & 422 & 386 & 372 & 362 & 385 & 362 \\ \hline
Average Cuts                   & -2.000 & -1.87 & -1.85 & -1.99 & -2.00 & -3.88 & -3.36 & -3.54 & -3.76 & -3.91 & -5.58 & -5.19 & -5.31 & -5.67 & -5.81 \\ \hline

\end{tabular}
\label{Table:t3}
\end{table*}

\begin{table*}[!htbp]
\centering
\renewcommand{\arraystretch}{1}
\renewcommand{\tabcolsep}{1.5mm}
\caption{Performance comparison of shot allocation methods applied to VQE simulations for LiH under noisy condition. In the VQE simulations, the energy calculated as the expectation value corresponds to the electronic ground-state energy. To obtain the total ground-state energy, the nuclear repulsion energy was added. All energy values are reported in Hartree units.}
\begin{tabular}{|c|c|c|c|c|c|c|c|c|c|c|}
\hline
 QPU Arch. & \multicolumn{5}{c|}{Heron} & \multicolumn{5}{c|}{Eagle} \\ \hline
\multirow{2}{*}{Application} & \multicolumn{5}{c|}{LiH VQE} & \multicolumn{5}{c|}{LiH VQE} \\ \cline{2-11} 
                             & Std. & lin\_fn & step\_fn & DDS & DDS\_M & Std. & lin\_fn & step\_fn & DDS & DDS\_M \\ \hline
$S_{avg}$                    & 1,024 & 89.73 
 & 100.21 & 532.97 & 616.26  & 1,024 & 76.43  & 99.25 & 554.69 & 569.45\\ \hline
Iterations                   & 689 & 697
 & 662 & 717 & 670 & 694 & 861 
 & 670 & 702 & 653 \\ \hline
Energy (E$_h$)                   & -7.86 
 & -7.64 & -7.62 & -7.86 & -7.87 & -7.87 
 & -7.67 & -7.46 & -7.86 & -7.86 \\ \hline

\end{tabular}
\label{Table:t4}
\end{table*}

We applied various shot allocation methods to VQA in a noisy environment and analyzed their impact on the total number of shots and accuracy. The qiskit-aer qasm simulator was employed to simulate noise, using the native gate sets and qubit errors of the ibm\_marrakesh and ibm\_yonsei processors, which are based on the Heron and Eagle architectures. For the VQA models, we utilized QAOA with SK models across various qubit numbers and VQE for LiH.

Figures \ref{fig:21} and \ref{fig:23} illustrate the shot count and cost per iteration for QAOA simulations with the SK model and VQE simulations for the LiH molecule, using the standard, linear\_fn, step\_fn, DDS, and DDS\_M shot allocation methods. In these simulations, the expectation value computed through the VQE circuit corresponds to the electronic ground state energy, with the total ground state energy obtained by adding the nuclear repulsion energy. Even in a noisy environment, DDS demonstrates good convergence, whereas linear\_fn and step\_fn fail to fully converge by the final iteration, displaying oscillatory behavior in their graphs. DDS\_M, which allows iterations with shot counts exceeding 1,024, shows improved convergence in the cost function compared to other methods.

Figures \ref{fig:22} and \ref{fig:24} illustrate the average cuts calculated at each iteration for QAOA simulations applied to the SK model and VQE simulations for LiH, comparing various shot allocation methods to the standard shot allocation approach. The results demonstrate that both linear\_fn and step\_fn exhibit slower convergence compared to DDS across all VQA models. Moreover, these methods fail to converge to a steady value and instead show oscillatory behavior throughout the iterations, persisting until the final iteration. In the QAOA SK model training process, significant fluctuations in the average cuts error are observed during earlier iterations, which makes the fluctuations in the final iteration less apparent. However, during the LiH VQE training process, these differences are more pronounced and clearly visible.

Table \ref{Table:t3} and Table \ref{Table:t4} present the final cost, total iterations (i.e., epochs), and the average number of shots ($S_{avg}$) per iteration for different shot allocation methods used in the training of each VQA model under noisy conditions. The linear\_fn and step\_fn methods exhibit a much smaller average number of shots per iteration compared to other shot allocation methods. However, when examining the final cost values, average cuts, and energy, it is evident that these values do not converge as well as those obtained using the standard method. In the case of DDS, the overall number of iterations is similar, but the average number of shots is reduced by half, with the final cost remaining almost the same. DDS\_M requires fewer average shots than the standard for small models, but more for the large 12-qubit SK QAOA models. Nevertheless, the final cost, average cuts, and energy are lower than the standard, indicating better accuracy.
\section{\label{sec:Appendix-G} Shot Reduction with Adjusting Optimizer}

Previous efforts to reduce shot counts in VQAs have introduced methods such as iCANS \cite{kubler2020adaptive}, Rosalin \cite{arrasmith2020operator}, gCANS \cite{gu2021adaptive}, weCANS \cite{ito2023latency}, and SantaQlaus \cite{ito2023santaqlaus}, which optimize the shot allocation process by modifying the optimizer within the VQA. Specifically, iCANS \cite{kubler2020adaptive} determines the sample count per iteration to maximize expected gain in the stochastic gradient descent process. While iCANS allows for individual variation of shot counts across gradient components, gCANS \cite{gu2021adaptive} applies a global optimization strategy to maximize expected gain for the entire gradient vector, and weCANS \cite{ito2023latency} focuses on maximizing gain per unit of time. Although integrating DDS with these optimizer-based schemes presents challenges, modifying the optimizers within VQAs makes it difficult to apply them together with other shot optimization methods. Future work could explore combining these approaches to improve shot efficiency in VQA.
\section{\label{sec:Appendix-H} Variance-Minimization-Based Shot Assignment}

Variance-minimization-based shot assignment strategies \cite{wecker2015progress, arrasmith2020operator, crawford2021efficient, mniszewski2021reduction, choi2022improving, zhang2023composite, zhu2024optimizing} aim to reduce total measurement variance by grouping commutable Hamiltonian terms into cliques, allowing simultaneous measurement of each clique. In this approach, the number of measurements for each clique, $N$, is optimized to minimize variance across all measurements. The VPSR (Variance-Preserving Shot Reduction) \cite{zhu2024optimizing} technique further refines this process by reducing the total number of shots while maintaining variance. VPSR initially measures $k$ shots for each clique to estimate its standard deviation and then uses the aggregated standard deviation data to effectively allocate an optimized shot count per clique. Although VPSR effectively reduces total shot usage while preserving variance, the initial $k$-shot per clique, followed by a standard deviation aggregation and shot allocation phase, introduces additional resource and computational overhead.

\end{document}